\documentclass[a4paper,11pt]{article}
\pdfoutput=1 

\usepackage{jcappub} 
\usepackage{afterpage}
\usepackage{graphics}
\usepackage[english]{babel}
\usepackage[utf8]{inputenc}
\newcommand{\planck}{\textit{\negthinspace Planck\/}}
\newcommand{\Planck}{\planck}
\newcommand{\dms}{\Delta^2_\mathrm{rms}}
\usepackage{color}

\title{\huge Power Spectra Based Planck Constraints on Compensated Isocurvature, and Forecasts for LiteBIRD and CORE Space Missions}

\author{Jussi V\"{a}liviita}

\affiliation{University of Helsinki, Department of Physics
and Helsinki Institute of Physics,\\
P.O. Box 64, FIN-00014 University of Helsinki, Finland}

\emailAdd{jussi.valiviita@helsinki.fi}

\abstract{Compensated isocurvature perturbations (CIP), where the primordial baryon and cold dark matter density perturbations cancel, do not cause total matter isocurvature perturbation. Consequently, at the linear order in the baryon density contrast $\Delta$, a mixture of CIP and the adiabatic mode leads to the same CMB spectra as the pure adiabatic mode. Only recently, Muñoz et al. showed that at the second order CIP leaves an imprint in the observable CMB by smoothing the power spectra in a similar manner as lensing. This causes a strong degeneracy between the CIP variance $\Delta_\mathrm{rms}^2\equiv\langle\Delta^2\rangle$ and the phenomenological lensing parameter $A_\mathrm{L}$. 
We study several combinations of the \Planck\ 2015 data and show that the measured lensing potential power spectrum $C_\ell^{\phi\phi}$ breaks the degeneracy. Nested sampling of the $\Lambda$CDM+$\dms$(+$A_\mathrm{L}$) model using the \Planck\ 2015 temperature, polarization, and lensing data gives $\dms=(6.9^{+3.0}_{-3.1})\times 10^{-3}$ at 68\% CL. A non-zero value is favoured at 2.3$\sigma$ (or without the polarization data at 2.8$\sigma$). CIP with $\dms\approx 7\times10^{-3}$ improves the bestfit $\chi^2$ by 3.6 compared to the adiabatic $\Lambda$CDM model. In contrast, although the temperature data favour $A_\mathrm{L}\simeq1.22$, allowing $A_\mathrm{L}\ne1$ does not improve the joint fit at all, since the lensing data disfavour $A_\mathrm{L}\ne1$. Indeed, CIP provides a rare example of a simple model, which is capable of reducing the \Planck\ lensing anomaly significantly and fitting well simultaneously the high (and low) multipole temperature and lensing data, as well as the polarization data. Finally, we derive forecasts for two future satellite missions (LiteBIRD proposal to JAXA/NASA and Exploring Cosmic Origins with CORE proposal to ESA's M5 call) and compare these to simulated \Planck\ data. Due to its coarse angular resolution, LiteBIRD is not able to improve the constraints on $\dms$  or $A_\mathrm{L}$, but CORE-M5 (almost) reaches the cosmic variance limit and improves the CIP constraint to $\dms < 0.6\,(1.4)\times10^{-3}$ at 68\,(95)\% CL, which is nine times better than the current trispectrum based upper bound and six times better than obtained from the simulated \Planck\ data. In addition, CORE-M5 will exquisitely distinguish between $\dms$ and $A_\mathrm{L}$. No matter whether CIP is allowed for or not, the uncertainty of the lensing parameter will be $\sigma(A_\mathrm{L})\approx0.012$, in the case where the simulated data are based on the adiabatic $\Lambda$CDM model with $A_\mathrm{L}=1$.}


\begin{document}

\begin{flushleft}
	\hfill		 HIP-2017-2/TH\\
\vspace{-20pt}
\end{flushleft}

\maketitle
  
\setcounter{tocdepth}{2}

\setcounter{secnumdepth}{3}

\section{Introduction}

Determination of the nature of primordial seed perturbations for structure formation and cosmic microwave background (CMB) anisotropies plays a crucial role in constraining inflationary models. The CMB and large scale structure observations indicate that the initial conditions of perturbations are predominantly adiabatic \cite{Pierpaoli:1999zj,Enqvist:2000hp,Enqvist:2001fu}, i.e., excited by the primordial curvature perturbation. The observations are consistent with a zero amplitude of the other possible initial modes, matter density isocurvature and neutrino density isocurvature, as well as, neutrino velocity isocurvature mode. Recently, the final (9th) data release \cite{Bennett:2012zja} of the Wilkinson Microwave Anisotropy probe (WMAP) and the second release \cite{Adam:2015rua} of \Planck\ satellite have set tight upper bounds on the fractional contribution of aforementioned modes to the primordial perturbations \cite{Savelainen:2013iwa,Ade:2015lrj}. However, those results do not constrain individually cold dark matter (CDM) density and baryon density isocurvature modes. This is due to the fact that the total matter density isocurvature perturbation can be small, if there is a (partial) cancellation between the CDM and baryon density isocurvature modes. We focus in this paper to an exact cancellation, i.e., study compensated isocurvature perturbations (CIP), where the total matter density isocurvature mode is zero, but CDM and baryon density perturbations may exceed even by orders of magnitude the primordial curvature perturbation, which anyway is responsible of producing the main observational features of the CMB angular power spectra.

To the linear order, CDM and baryon density isocurvature are indistinguishable in the CMB or matter power spectrum \cite{Gordon:2002gv}.
Therefore it was believed that in order to constrain CIP one had to go to the trispectrum level \cite{Grin:2013uya} or resort to the observations of the distribution of neutral hydrogen between redshifts
30 to 300 using 21\,cm absorption lines \cite{Gordon:2009wx} and the constraints on the spatial variation of the baryon fraction in galaxy clusters \cite{Holder:2009gd,Grin:2011tf,Grin:2013uya}.  However, in \cite{Munoz:2015fdv}  it was shown that CIP leaves an observable effect even to the CMB power spectra, if one goes to the second order in the CIP amplitude $\Delta_\mathrm{rms}$, which we introduce in the next section in equations \eqref{eq:CIPcond3}--\eqref{eq:obscompensatedisoc}. This power spectra based analysis is much more straightforward and transparent than the trispectrum analysis. The two methods were shown to lead to constraints of the same order of magnitude by \cite{Munoz:2015fdv}.

\noindent We extend the analysis of \cite{Munoz:2015fdv} in several ways. 
{\small\begin{itemize}
\vspace{-1.0mm}\item[(1)] We replace the Fisher matrix analysis of the \Planck\ 2015 CMB data by full nested sampling and evaluate the effect of CIP at each point in parameter space instead of precalculating it only for the bestfit \Planck\ $\Lambda$CDM cosmology.
\vspace{-1.5mm}\item[(2)] This allows us to study in more detail the interesting degeneracy between the CIP variance $\dms$ and the phenomenological lensing parameter $A_\mathrm{L}$ that scales the power spectrum of the lensing potential by a scale-independent amplitude compared to the standard $\Lambda$CDM model where $A_\mathrm{L}=1$. A smaller value reduces the smoothing caused by lensing at high multipoles of temperature and E-mode polarization power spectra. A larger value increases the smoothing effect. A positive $\dms$ leads to a very similar observational effect as $A_\mathrm{L}>1$, to be discussed in section \ref{sec:degeneracy}. It turns out that the \Planck\ lensing anomaly, i.e., the high multipole temperature data favouring $A_\mathrm{L} \sim 1.22$ within the $\Lambda$CDM cosmology could be mitigated by a positive $\dms$. However, there are many other one-parameter extensions to the adiabatic $\Lambda$CDM model that also help fitting the high-$\ell$ temperature data. The actual problem is that none of commonly studied one-parameter extensions significantly help fitting simultaneously the \Planck\ high-$\ell$ (and low-$\ell$) temperature and lensing data \cite{Ade:2013zuv,Ade:2015xua,Addison:2016}. One can ``easily'' devise models that fit one of these datasets (derived from the same measurements), but then the other dataset is typically fit even worse than by the $\Lambda$CDM model.
\vspace{-1.5mm}\item[(3)] Ref.~\cite{Munoz:2015fdv} ignored the lensing data. We include these data in the end of our \Planck\ analysis, and find (see table \ref{tab:RealPlanckChi2}) that CIP with $\dms \approx 7\times10^{-3}$ (and $A_\mathrm{L}=1$) fits well both the \Planck\ temperature and polarization data, as well as simultaneously providing an excellent fit to the lensing data.
\end{itemize}}

After establishing the current \Planck\ constraints in section \ref{sec:planck} we proceed to the forecasts for future satellite missions (CORE-M5 and LiteBIRD) in section \ref{sec:future}.
CORE-M5 is a proposal submitted in response to European Space Agency's (ESA) call for medium size space missions for launch in 2029--2030. LiteBIRD, or Lite satellite for the study of B-mode polarization and Inflation from cosmic microwave background Radiation Detection, is a proposal to the  Japan Aerospace Exploration Agency (JAXA) with the launch year in the early 2020s \cite{Matsumura:2013aja,Matsumura:2016sri,Grenoble}. According to \cite{Grenoble} a similar proposal has been submitted also to National Aeronautics and Space Administration (NASA).

A detailed account on various aspects of the CORE-M5 proposal can be found in the ``Exploring Cosmic Origins with CORE'' publication series \cite{ecoMission,ecoInstrument,DiValentino:2016foa,Finelli:2016cyd,ecoCompSep,ecoCluster,DeZotti:2016qfg}. While \cite{DiValentino:2016foa} deals with the forecasts for determination of cosmological parameters in the standard $\Lambda$CDM model and (typically one-parameter) extensions to it, Ref.~\cite{Finelli:2016cyd} focuses on constraining inflationary models. It includes detailed forecasts for CORE-M5, down/upscaled CORE-like missions, and LiteBIRD for the determination accuracy of the initial conditions of perturbations by studying a mixture of adiabatic and CDM isocurvature perturbations by introducing one or three extra ``non-adiabaticicy'' parameters compared to the standard adiabatic $\Lambda$CDM model. CIP is left for this separate paper. 

Ref.~\cite{Finelli:2016cyd} quotes one of our results: $\dms < 0.0019$ at 95\% CL for the adiabatic $\Lambda$CDM fiducial cosmology using the ``minimal'' set of CORE-M5 data, i.e., only the temperature, E-mode polarization, and their cross-correlation spectra. In this paper we go beyond these data and improve the constraint by including CORE-M5 lensing potential data, and in addition show that CORE-M5 will virtually break the degeneracy between $\dms$ and $A_\mathrm{L}$. As a side product of our analysis we obtain forecasts for the determination of $A_\mathrm{L}$ that is not discussed in \cite{Finelli:2016cyd,DiValentino:2016foa}. For comparison, we apply the same analysis pipeline to the LiteBIRD and simulated \Planck\ data, and to an ideal cosmic variance limited experiment, where instrumental noise is zero.

\section{General initial conditions for perturbations}
\label{sec:cip}

\subsection{Adiabatic and isocurvature modes}

Usually, in the cosmological analysis, adiabatic initial conditions are assumed for primordial perturbations, deep at the radiation dominated epoch of the evolution of Universe. This means that the seeds for the observed structure (galaxies, galaxy clusters) and the small anisotropies of the CMB radiation can be described by a small primordial (comoving) curvature perturbation $\mathcal{R}$. Further, this implies that the entropy density is spatially constant, i.e, the number densities of different particle species, such as radiation (photons and neutrinos) and matter (baryons and CDM) fluctuate in space hand in hand: wherever the number density of one species is larger than the average, there also the others have an over density, and vice versa. This assumption can be motivated by the simplest single-field slow-roll inflationary models. They have only one degree of freedom for perturbations, namely the spatial (quantum) fluctuations of the inflaton field, which causes tiny spatial curvature perturbations that are stretched to classical ones due to the rapid expansion of the Universe during inflation. After inflation, in reheating, all the primordial inhomogeneities of the Universe are created from this single quantity, which excites the adiabatic initial perturbation mode.

However, more complicated models have more degrees of freedom and may lead to spatial entropy perturbations. Typical examples are multi-field inflationary models. In the two-field case one field could produce the radiation and another field the matter. This can introduce spatial variation to the relative primordial number densities: say $\mathcal{S}_{mr} \equiv \delta(n_m/n_r)\,/\,(n_m/n_r)$ is not identically zero (as it would be in the adiabatic case). Here $n_m$ and $n_r$ are the number densities of matter and radiation particles, respectively. The quantity $\mathcal{S}_{mr}$ is an entropy, i.e., isocurvature perturbation between matter and radiation and it can be written as
\begin{equation}
\mathcal{S}_{mr} = \frac{\delta n_m}{n_m} - \frac{\delta n_r}{n_r} = \frac{\delta\rho_m}{\rho_m} - \frac{3}{4}\frac{\delta\rho_r}{\rho_r} = \frac{\delta_m}{1+w_m} - \frac{\delta_r}{1+w_r}\,,
\end{equation}
where $\rho_m$ and $\rho_r$ are the average radiation and matter energy densities, $\delta\rho_m$ and $\delta\rho_r$ their perturbations, and $\delta_m\equiv\delta\rho_m/\rho_m$ and  $\delta_r\equiv\delta\rho_r/\rho_r$. The equation of state parameters are $w_m\equiv p_m/\rho_m=0$ for the matter and $w_r\equiv p_r/\rho_r=1/3$ for the radiation. In a similar manner we can write other relative entropy perturbations between two different species $i$ and $j$: $\mathcal{S}_{ij} \equiv -3H(\delta\rho_i/\dot\rho_i - \delta\rho_j/\dot\rho_j ) = \delta_i / (1+w_i) - \delta_j / (1+w_j)$, where we used the continuity equation $\dot\rho_i = -3H(1+w_i)\rho_i$. We can further write
\begin{equation}
\mathcal{S}_{mr} = \frac{\rho_c}{\rho_m} \mathcal{S}_{cr} +  \frac{\rho_b}{\rho_m} \mathcal{S}_{br} =  \frac{\Omega_c}{\Omega_m} \mathcal{S}_{cr} +  \frac{\Omega_b}{\Omega_m} \mathcal{S}_{br}\,,
\end{equation}
where $\rho_c$ and $\rho_b$ are the average CDM and baryon energy densities, and $\Omega$s are the density parameters with respect to the critical density. If there is no neutrino density isocurvature perturbation between neutrinos ($\nu$) and photons ($\gamma$), i.e., $\mathcal{S}_{\nu\gamma} = 0$, then $\mathcal{S}_{cr} = \mathcal{S}_{c\nu} = \mathcal{S}_{c\gamma}$ and  $\mathcal{S}_{br} = \mathcal{S}_{b\nu} = \mathcal{S}_{b\gamma}$. So, we have
\begin{equation}
\mathcal{S}_{mr} = \mathcal{S}_{m\gamma}  = \frac{\Omega_c}{\Omega_m} \mathcal{S}_{c\gamma} +  \frac{\Omega_b}{\Omega_m} \mathcal{S}_{b\gamma}\,,
\label{eq:matteriso}
\end{equation}
where $\mathcal{S}_{c\gamma}$ is CDM density isocurvature perturbation and $\mathcal{S}_{b\gamma}$ is baryon density isocurvature perturbation.

The most general (growing mode) initial conditions are an arbitrarily correlated mixture \cite{Langlois:1999dw,Langlois:2000ar,Gordon:2000hv,Amendola:2001ni} of the curvature perturbation $\mathcal{R}$ (the adiabatic mode), the CDM density isocurvature mode $\mathcal{S}_{c\gamma}$, the baryon density isocurvature mode $\mathcal{S}_{b\gamma}$, the neutrino density isocurvature mode $\mathcal{S}_{\nu\gamma}$, and a neutrino velocity isocurvature mode  \cite{Bucher:1999re,Bucher:2000hy}. However, by today no theoretically compelling mechanism for exciting the last one has been presented. The three density isocurvature modes can be naturally excited during multi-field inflation, but thermalization after the end of inflation may erase the isocurvature signature by the primordial time \cite{Mollerach:1989hu,Weinberg:2004kf,Beltran:2005gr}. Therefore, a non-detection of isocurvature does not rule out multi-field inflation, but as the single-field slow-roll inflation (with a canonical kinetic term) can only excite the adiabatic mode, any detection of isocurvature would rule out these simplest inflationary models. Thus the study of the nature of initial conditions is a crucial part of constraining the inflationary models.

\subsection{Compensated isocurvature perturbations (CIP)}

The adiabatic, matter density isocurvature, neutrino density isocurvature and neutrino velocity isocurvature mode each leave distinct imprints in the phases and relative amplitudes of the peaks and dips of the CMB temperature and polarization angular power spectra (see e.g. \cite{Savelainen:2013iwa,Planck:2013jfk,Ade:2015lrj}), as well as the matter power spectrum and the phase of baryon acoustic oscillations (BAO) \cite{Valiviita:2012ub}. 
However, at the linear order, fixed non-zero values $\mathcal{S}_{b\gamma}=S$ or $\mathcal{S}_{c\gamma}=(\Omega_b/\Omega_c)S$  leave identical imprint at observable scales \cite{Gordon:2002gv}. Therefore, the individual amplitudes of $\mathcal{S}_{c\gamma}$ and $\mathcal{S}_{b\gamma}$ are rather unconstrained as long as the amplitude of the matter density isocurvature mode $\mathcal{S}_{m\gamma}$ stays much smaller than $\mathcal{R}$. In an extreme case, where
\begin{equation}
\mathcal{S}_{c\gamma} = -\frac{\Omega_b}{\Omega_c} \mathcal{S}_{b\gamma}\,,
\label{eq:CIPcond1}
\end{equation}
the matter density isocurvature vanishes according to \eqref{eq:matteriso} and there is no linear order isocurvature signal in the CMB power spectra. Perturbations obeying \eqref{eq:CIPcond1} are called compensated CDM and baryon isocurvature perturbations. At the primordial time the photon density dominates over the CDM and baryon density, which means that $\delta_\gamma$ is negligible and \eqref{eq:CIPcond1} can be written as $\delta_c \approx -(\Omega_b/\Omega_c)\delta_b$, which simplifies to
\begin{equation}
\delta\rho_c(t_\mathrm{primordial},\mathbf{x}) \approx -\delta\rho_b(t_\mathrm{primordial},\mathbf{x})\,.
\label{eq:CIPcond2}
\end{equation}
\afterpage{
\begin{figure}
\centering
\includegraphics[width=0.6560\textwidth]{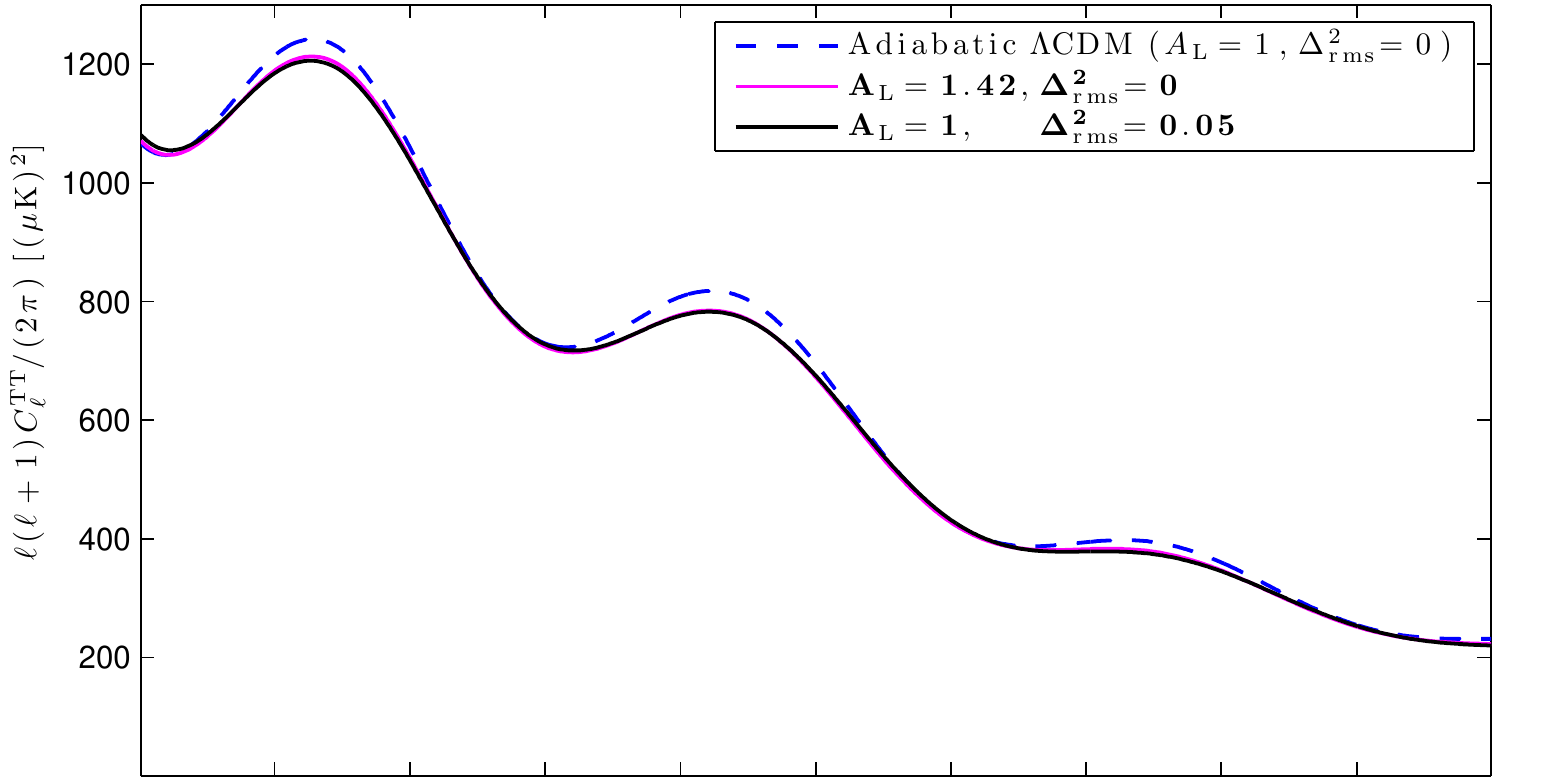}\\
\includegraphics[width=0.6488\textwidth]{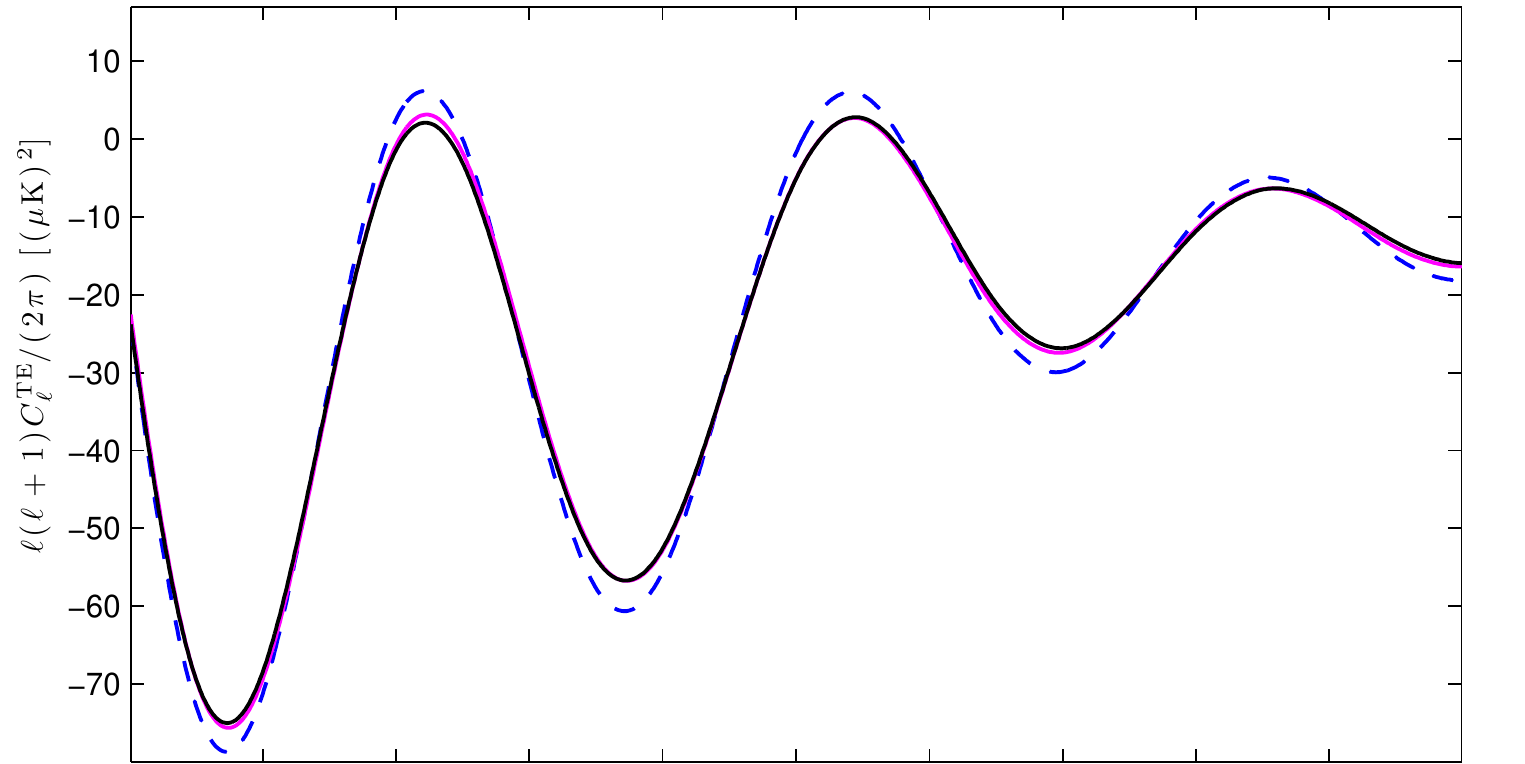}\\
\includegraphics[width=0.6400\textwidth]{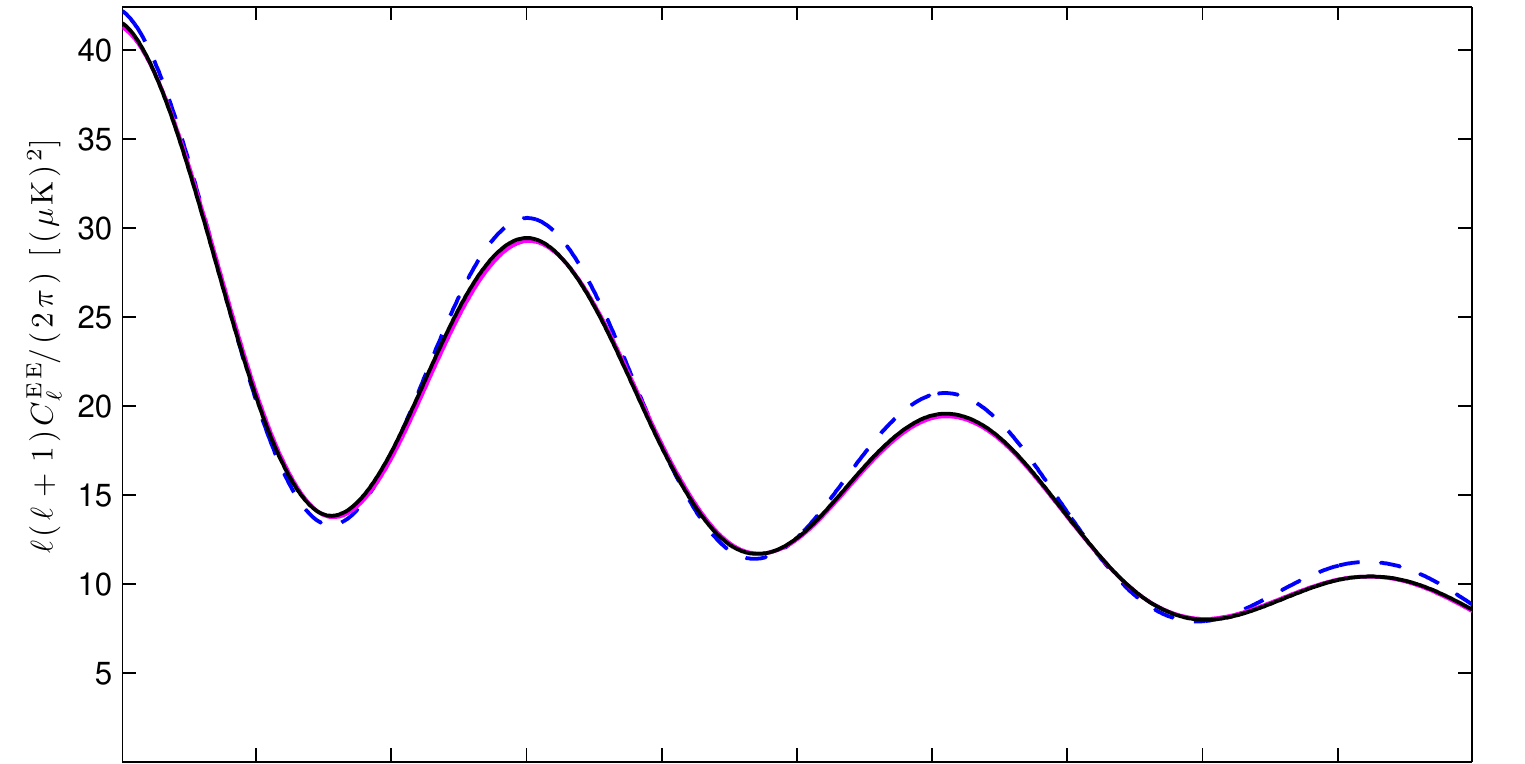}\\
\includegraphics[width=0.662\textwidth]{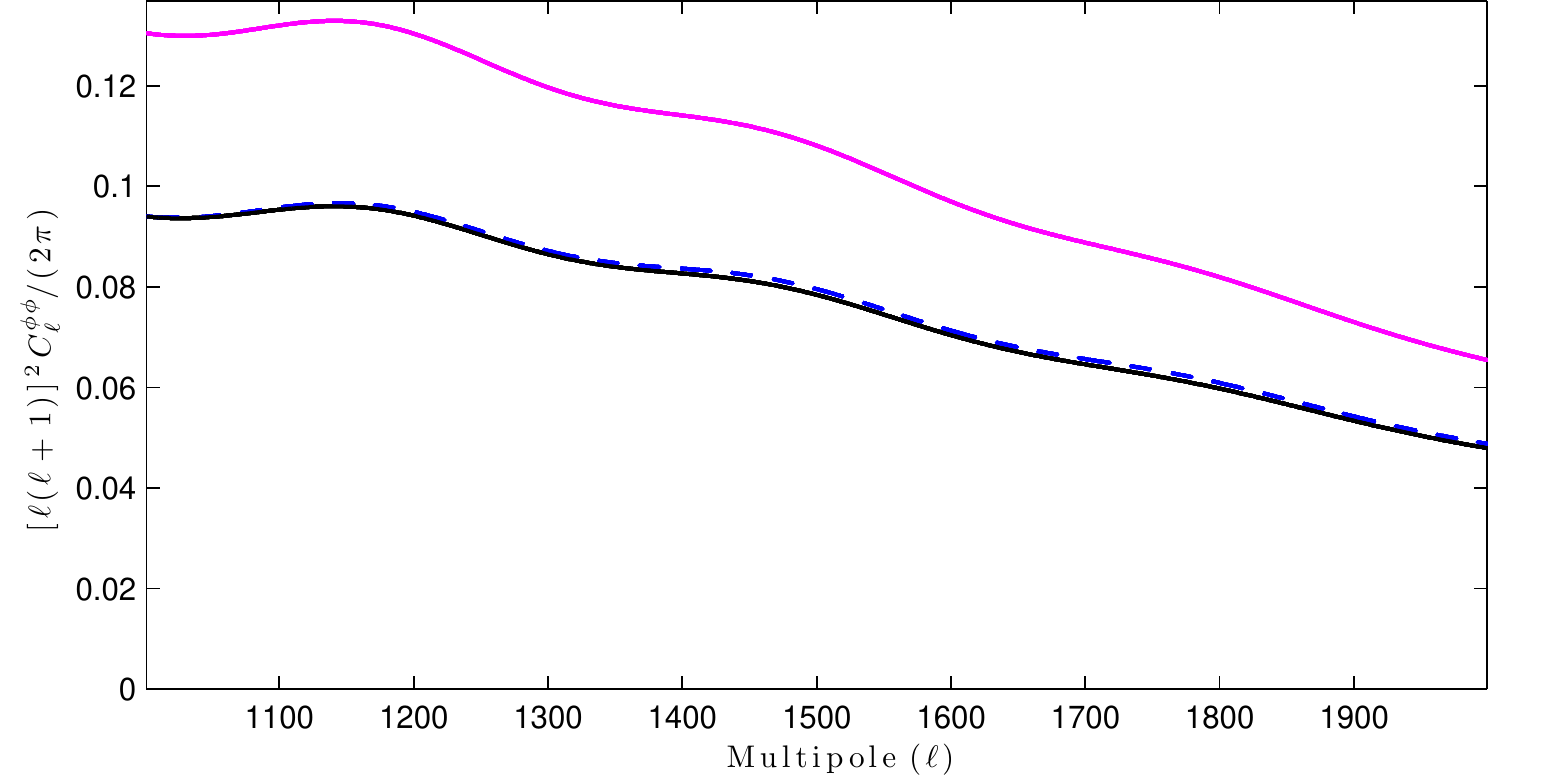}\\
\vspace{-2.5mm}
\caption{A comparison of the angular spectra of the adiabatic $\Lambda$CDM model (blue dashed lines), adiabatic $\Lambda$CDM+$A_\mathrm{L}=1.42$ (magenta solid lines), and compensated isocurvature model with $\dms=0.05$ (black solid lines).  The cosmological parameters of these example models are given in table \ref{tab:fiducial}.
\label{fig:Cell_AL_LCDM}}
\end{figure}
\begin{table}
\setlength{\tabcolsep}{1.15mm}
\footnotesize
\begin{tabular}{|l|rrrrrrrrr|}
\hline
 & $\omega_\mathrm{b}$ & $\omega_\mathrm{c}$ & $\tau$ & $H_0$ & $10^{10}A_\mathrm{S}$ & $n_\mathrm{S}$ & ($\Omega_\Lambda$) &  $\dms$  & $A_\mathrm{L}$ \\
\hline
 \textcolor{blue}{\bf Fiducial/example abiab.\ $\mathbf{\Lambda}$CDM}                                 &  0.02214 &  0.1206 & 0.0581 &  66.89 &  21.179 & 0.9625 & (0.681)  &  \textcolor{blue}{\bf 0} & \textcolor{blue}{\bf 1.00} \\
\textcolor{magenta}{\bf Example adiab.\ $\mathbf{\Lambda}$CDM+$\mathbf{A_\mathrm{L}}$}     &  0.02160  &  0.1194  & 0.0559 &  66.39 &  21.068 & 0.9321 & (0.680)  &   \textcolor{magenta}{\bf 0}            & \textcolor{magenta}{\bf 1.42} \\
 {\bf Example CIP}                               &  0.02214 &  0.1206 & 0.0581 &  66.89 &  21.179 & 0.9625 & (0.681)  & {\bf 0.05} & {\bf 1.00} \\
\hline
\end{tabular}
\caption{Parameters of the three example models shown in figure \ref{fig:Cell_AL_LCDM}. The first one is also used to create simulated adiabatic $\Lambda$CDM data in section \ref{sec:future}. $\Omega_\Lambda$ is calculated from the other parameters. \label{tab:fiducial}}
\end{table}}

In principle, between the primordial time and last scattering baryons behave differently from dark matter at small scales and CIP is expected to modify the angular power spectrum with respect to the pure adiabatic one, but these scales correspond to multipoles $\ell > 10^5$--$10^6$ \cite{Grin:2011tf}. At much larger scales the primordial set-up of \eqref{eq:CIPcond2} is preserved and CIP can be described as a small anisotropy of baryon and CDM density \cite{Munoz:2015fdv,Heinrich:2016gqe}
\begin{eqnarray}
\label{eq:CIPcond3}
\Omega_\mathrm{b}(\hat n) & = & [1+\Delta(\hat n)]\bar\Omega_\mathrm{b}\,,\\
 \Omega_\mathrm{c}(\hat n) & = & \bar\Omega_\mathrm{c} -\Delta(\hat n)\bar\Omega_\mathrm{b}\,.
\end{eqnarray}
Here the overbar denotes the average over the whole sky and $\Delta(\hat n)$ a small perturbation about this average in the direction $\hat n$.
According to Ref.~\cite{Munoz:2015fdv}, $\Delta$ can be treated as a Gaussian random variable, which has a zero mean and variance $\Delta^2_\mathrm{rms}$.
The observed angular power is then an average over the values of $\Delta$
\begin{equation}
C_\ell^\mathrm{obs}(\bar\Omega_\mathrm{b},\bar\Omega_\mathrm{c},\tau,H_0,n_\mathrm{S},A_\mathrm{S}) = \frac{1}{\sqrt{2\pi\Delta^2_\mathrm{rms}}} \int  \!\!C_\ell\big(\Omega_\mathrm{b}(\Delta),\Omega_\mathrm{c}(\Delta),\tau,H_0,n_\mathrm{S},A_\mathrm{S}\big) \, e^{-\Delta^2/(2\Delta^2_\mathrm{rms})} d\Delta\,,
\label{eq:obscompensatedisoc1}
\end{equation}
where $\Omega_\mathrm{b}(\Delta) =  (1+\Delta)\bar\Omega_\mathrm{b}$ and $\Omega_\mathrm{c}(\Delta) = \bar\Omega_\mathrm{c} -\bar\Omega_\mathrm{b}\Delta$, and $C_\ell$ is calculated by assuming the adiabatic initial conditions.
This averaging over slightly different values of $\Omega_\mathrm{b}$ (and $\Omega_\mathrm{c}$) leads to a lensing-like effect in the power spectra \cite{Munoz:2015fdv}, which we discuss in detail in the following sections.
Substituting into \eqref{eq:obscompensatedisoc1} the Taylor expansion of $C_\ell$ about $\Delta=0$, one finds
\begin{equation}
C_\ell^\mathrm{obs} \approx C_\ell|_{\Delta=0} + \frac{1}{2} \frac{d^2 C_\ell}{d\Delta^2}\Big|_{\Delta=0}\Delta^2_\mathrm{rms}\,.
\label{eq:obscompensatedisoc}
\end{equation}

In \cite{Munoz:2015fdv} Fisher matrix constraints on $\Delta^2_\mathrm{rms}$ were presented for the \Planck\ data and for an ideal cosmic variance limited case with $\ell_\mathrm{max}=2500$ in a model where CIP and the adiabatic mode are uncorrelated. The second derivative appearing in \eqref{eq:obscompensatedisoc} was evaluated at a single pre-selected point in parameter space, namely the \Planck\ bestfit $\Lambda$CDM model. Here we use the full \texttt{MultiNest} \cite{Feroz:2007kg,Feroz:2008xx,Feroz:2013hea} nested sampling (together with a slightly modified version of \texttt{CosmoMC} \cite{Lewis:2002ah,Lewis:2013hha} and \texttt{CAMB} \cite{Lewis:1999bs,Howlett:2012mh}) and evaluate the second derivative numerically at each point of the parameter space. In principle, this is a more accurate method, since the bestfit region in the CIP model will differ from the bestfit region of the adiabatic $\Lambda$CDM model. However, we have verified that in practice the difference in the constraints on $\dms$ is small.

\section{Degeneracy Between CIP and Lensing Parameter $A_\mathrm{L}$}
\label{sec:degeneracy}

Before proceeding to the \Planck\ constraints we discuss the effect of CIP (a non-zero $\dms$) and the phenomenological lensing parameter $A_\mathrm{L}$ on the observable CMB and lensing power spectra. In figure \ref{fig:Cell_AL_LCDM} we show the angular power spectra of three different models zooming to the multipole range $\ell=1000$--$2000$.
(The values of cosmological parameters for these models are reported in table \ref{tab:fiducial}.) The blue dashed lines show the adiabatic $\Lambda$CDM model.
The solid black lines are obtained with the same parameters in a CIP model where $\dms=0.05$. We choose this relatively large value in order to make the effect of CIP easily visible. Solid magenta lines are for the adiabatic $\Lambda$CDM model with the value of $A_\mathrm{L}$ (and other parameters) chosen to lead to a close match with the CIP model.
From the first three panels it is obvious that a non-zero $\dms$ may lead to a very similar TT, TE, and EE angular power spectra as the adiabatic model with exaggerated lensing (by $A_\mathrm{L} = 1.42$ in this case). Due to this $\dms$--$A_\mathrm{L}$ degeneracy, it is difficult to tell whether the TT,TE,EE data have $A_\mathrm{L}=1$ and a positive $\dms$, or a zero $\dms$ and $A_\mathrm{L}>1$, or even a positive $\dms$ but $A_\mathrm{L}<1$.  However, the last panel indicates that the  power spectrum of the lensing potential, $C_\ell^{\phi\phi}$, may be used to break this degeneracy, since $A_\mathrm{L}$ simply multiplies the ordinary $C_\ell^{\phi\phi}$, whereas a non-zero $\dms$ has a negligible effect on $C_\ell^{\phi\phi}$.

\begin{figure}
\centering
\includegraphics{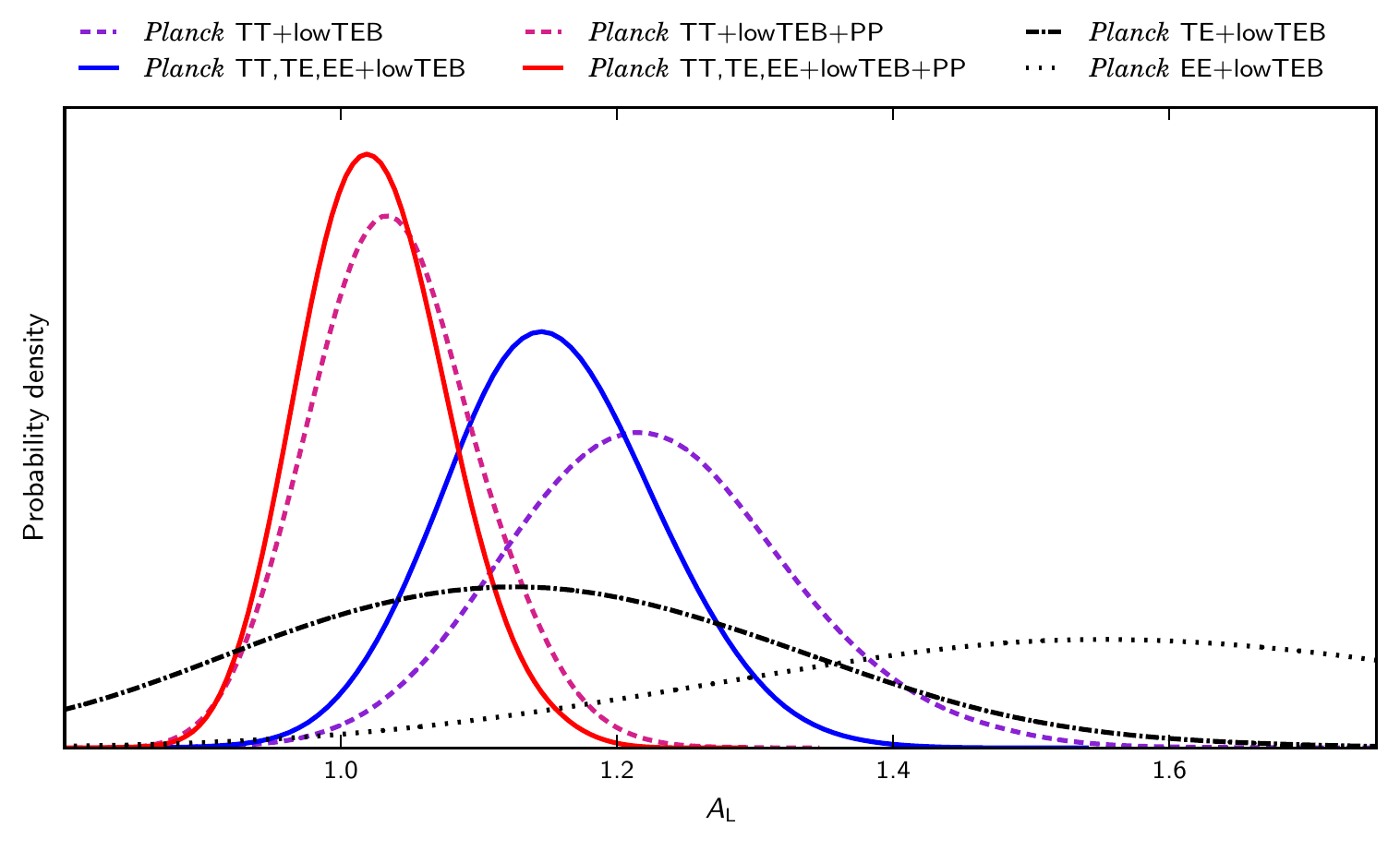}\\
\vspace{-5.5mm}
\caption{
The posterior probability density of the phenomenological lensing amplitude $A_\mathrm{L}$ with \Planck\ 2015 data using high-$\ell$ temperature (TT) and low-$\ell$ temperature and polarization data (lowTEB),  high-$\ell$ temperature and polarization data (TT,TE,EE) and lowTEB, the previous ones supplemented with the lensing potential reconstruction data (PP), and finally using only the polarization part of the high-$\ell$ data, either TE or EE, plus lowTEB.  The fitted model is $\Lambda$CDM+$A_\mathrm{L}$.
\label{fig:planckAL}}
\end{figure}

\section{Constraints from the \Planck\ 2015 Data}
\label{sec:planck}

For the current CMB power spectra based constraints we use the \Planck\ 2015 data release and various combinations of likelihood codes available in the Planck Legacy Archive (PLA)  \cite{pla} in a similar manner as done in \cite{Ade:2015xua,Ade:2015lrj}. At low multipoles ($\ell < 30$) we use always the temperature and polarization pixel based likelihood \texttt{lowl\_SMW\_70\_dx11d\_2014\_10\_03\_v5c\_Ap.clik} labelled as lowTEB. At high multipoles ($\ell \ge 30$) we employ either the temperature only likelihood \texttt{plik\_dx11dr2\_HM\_v18\_TT.clik} labelled as TT, or the temperature, polarization E-mode, and their cross-correlation likelihood \texttt{plik\_dx11dr2\_HM\_v18\_TTTEEE.clik} labelled as TT,TE,EE. We may also supplement either of these by the power spectrum of the lensing potential $[\ell(\ell+1)]^2C_\ell^{\phi\phi}/(2\pi)$ in the conservative range $40 \le \ell \le 400$ measured by \Planck, \texttt{smica\_g30\_ftl\_full\_pp.dataset}, labelled as PP. The temperature and polarization likelihoods are described in \cite{Aghanim:2015xee} and the lensing likelihood in \cite{Ade:2015zua}.

\subsection{Lensing amplitude in the $\Lambda$CDM+$A_\mathrm{L}$ model}

When using the real Planck data we should keep in mind the ``lensing anomaly''. In particular, the high-$\ell$ temperature power spectrum suggests that the amplitude of the lensing potential should be multiplied by $A_\mathrm{L} \approx 1.22$ with respect to the prediction of the standard $\Lambda$CDM model where $A_\mathrm{L}=1$ \cite{Ade:2015xua,Addison:2016,Couchot:2015eea}. (With \Planck\ TT+lowTEB, $A_\mathrm{L}$ is $2.2\sigma$ away from $1$.) This means that the high-$\ell$ peaks and droughts are more smoothed in the data than predicted by the $\Lambda$CDM model. Since the compensated isocurvature leads to a lensing-like effect, there will be a strong degeneracy (indeed a negative correlation) between $A_\mathrm{L}$ and $\dms$ as noticed in \cite{Munoz:2015fdv}. If $A_\mathrm{L}$ is fixed to unity, then the \Planck\ TT data will inevitably favour a non-zero $\dms$, which makes the interpretation of the constraints cumbersome. We start by reproducing some \Planck\ results from publicly available MCMC chains in PLA. From figure \ref{fig:planckAL} we notice that the temperature--E-mode polarization cross-correlation (TE) favours a slightly smaller $A_\mathrm{L}$ than TT, whereas EE autocorrelation favours even higher values\footnote{Note that, as discussed in \cite{Ade:2015xua}, the high-$\ell$ EE part of \Planck\ likelihoods is not very stable with respect to $A_\mathrm{L}$: the baseline \texttt{Plik} EE likelihood leads to $A_\mathrm{L}\sim 1.54$ as seen in figure \ref{fig:planckAL}, whereas the alternative \texttt{CamSpec} would give $A_\mathrm{L}\sim 1.19$. Both these EE likelihoods constrain $A_\mathrm{L}$ very weakly, $\sigma(A_\mathrm{L}) \sim 0.2$--$0.3$.} than TT. However, the weight of EE in the combined TT,TE,EE fit is so small that (due to TE) the peak of the posterior with TT,TE,EE ends up being at a slightly smaller value of $A_\mathrm{L}$ than with TT alone. What turns out important for us is that the lensing potential reconstruction data (PP) favour so much smaller values of $A_\mathrm{L}$ than TT or TT,TE,EE that with PP the posterior peaks at $A_\mathrm{L} \approx 1$. (See also the last column of Table~\ref{tab:RealPlanck}.) From this we deduce that, when determining $\dms$, the results should not depend on whether we vary $A_\mathrm{L}$ or not, if we include the PP data in the analysis. 
Without PP the dependence on $A_\mathrm{L}$ is expected to be very strong.
\begin{table}
\setlength{\tabcolsep}{0.40mm}
\footnotesize
\begin{tabular}{|l|rr|rr||rr||rr|}
\hline
 \raisebox{11pt} {}Model & \multicolumn{4}{c||}{$\Lambda$CDM+$A_\mathrm{L}$+$\dms$} &  \multicolumn{2}{c||}{$\Lambda$CDM+$\dms$} & \multicolumn{2}{c|}{$\Lambda$CDM+$A_\mathrm{L}$}\\
  \raisebox{10pt} {}    & \multicolumn{4}{c||}{(free $A_\mathrm{L}$ and $\dms$)} &  \multicolumn{2}{c||}{($A_\mathrm{L}=1$, free $\dms$)} & \multicolumn{2}{c|}{($\dms=0$, free $A_\mathrm{L}$)} \\ %
\hline
 \raisebox{11pt} {}Parameter    &  \multicolumn{2}{c|}{$1000\dms$} &  \multicolumn{2}{c||}{$A_\mathrm{L}$} &  \multicolumn{2}{c||}{$1000\dms$} &  \multicolumn{2}{c|}{$A_\mathrm{L}$}\\
\hline
 \Planck\ TT+lowTEB &     11.36  & $[     0.00;     14.37]$ &      1.06 & $[     0.94;      1.20]$ &     13.12 & $[     7.69;     18.38]$ &      1.22 & $[     1.12;      1.32]$  \\
 \Planck\ TT,TE,EE+lowTEB &      4.57 & $[     0.00;      \phantom{1}5.70]$ &      1.08 & $[     1.01;      1.15]$ &      6.22 & $[     2.12;       \phantom{1}9.10]$ &      1.15 & $[     1.07;      1.23]$  \\ 
 \Planck\ TT+lowTEB+PP &     11.03 & $[     7.08;     15.19]$ &      1.05 & $[     0.98;      1.10]$ &     10.68 & $[     6.64;     14.63]$ &      1.04 & $[     0.97;      1.09]$  \\
 \Planck\ TT,TE,EE+lowTEB+PP &      6.88 & $[     3.68;       \phantom{1}9.80]$ &      1.00 & $[     0.96;      1.05]$ &      6.87 & $[     3.79;       \phantom{1}9.85]$ &      1.02 & $[     0.97;      1.07]$  \\
\hline
\end{tabular}
\caption{The posterior mean value and 68\% CL interval for $1000\dms$ and $A_\mathrm{L}$ in three different models (first $\Lambda$CDM+$A_\mathrm{L}$+$\dms$, second  $\Lambda$CDM+$\dms$, third  $\Lambda$CDM+$A_\mathrm{L}$) using four different combinations of \Planck\ data. All $\Lambda$CDM and \Planck\ nuisance parameters are varied and finally marginalized over. The third case is reproduced from publicly available \Planck\ 2015 MCMC chains, while the first two cases are produced by our \texttt{MultiNest} runs. The first two cases are also shown in figure \ref{fig:planckDeltaAL1d}.
\label{tab:RealPlanck}}
\end{table}

\subsection{Compensated isocurvature in $\Lambda$CDM+$A_\mathrm{L}$+$\dms$ and $\Lambda$CDM+$\dms$ models}

Now we are ready for the new results. We vary the ordinary six $\Lambda$CDM parameters and $A_\mathrm{L}$ whenever indicated, 15 (with TT) or 27 (with TT,TE,EE) nuisance parameters of the \Planck\ likelihoods, and $\dms$, allowing for only non-negative values for the last one. 
{\begin{figure}
\centering
\includegraphics{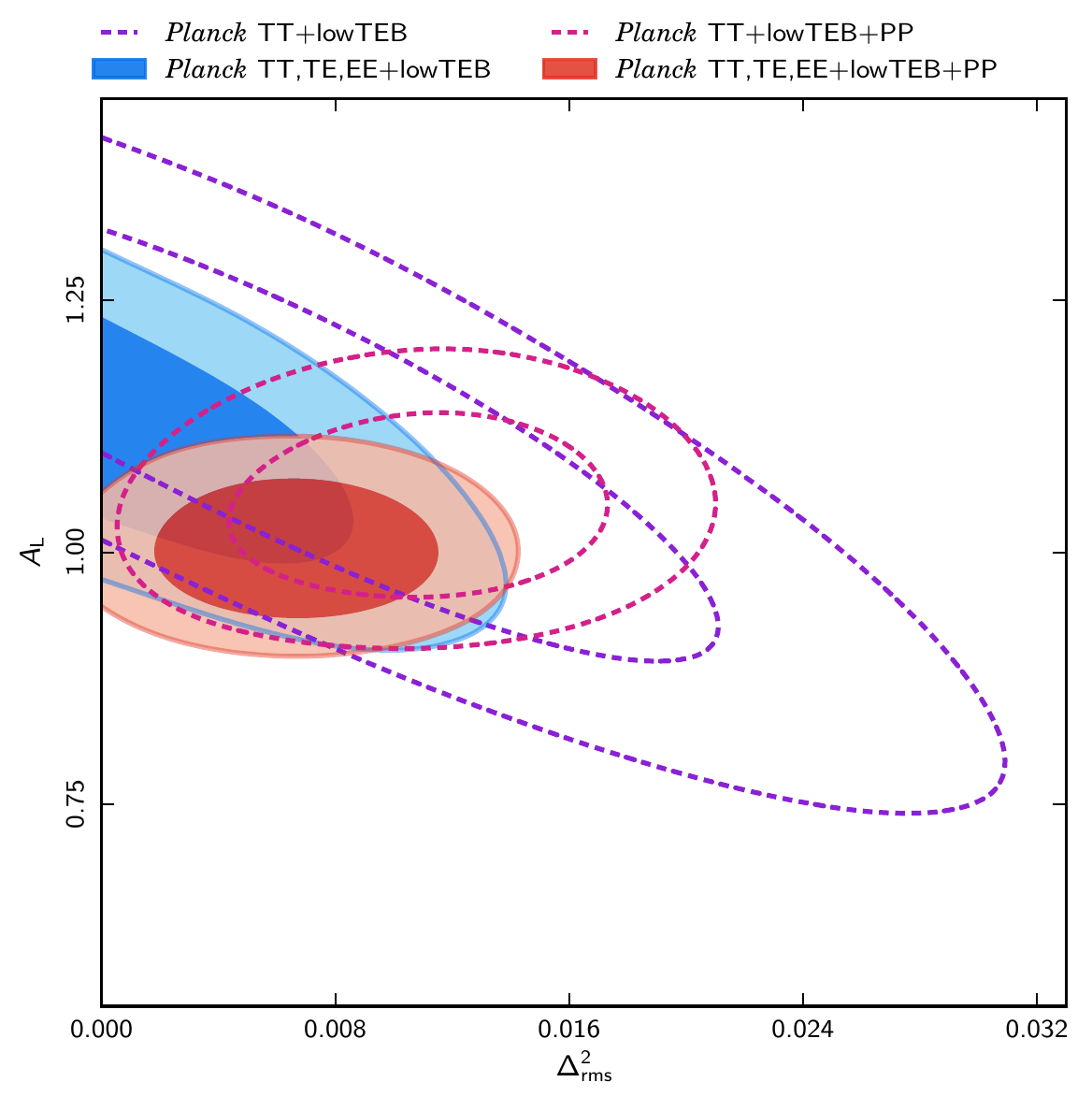}\\
\vspace{-5.5mm}
\caption{
2d marginalized 68\% and 95\% CL posterior regions of the phenomenological lensing amplitude $A_\mathrm{L}$ and the variance of compensated isocurvature amplitude $\dms$ with \Planck\ 2015 data using high-$\ell$ temperature (TT) data and low-$\ell$ temperature and polarization data (lowTEB),  high-$\ell$ temperature and polarization data (TT,TE,EE) and lowTEB, and the previous ones supplemented with the lensing potential reconstruction data (PP).  
\label{fig:planckDeltaAL2d}}
\end{figure}}
Figure \ref{fig:planckDeltaAL2d} indicates that with \Planck\ TT+lowTEB (TT,TE,EE+lowTEB) data $A_\mathrm{L} = 1$ becomes acceptable if $\dms \approx 0.016$ ($\dms \approx 0.010$).
Adding the lensing data has a dramatic effect: the best-fitting models are provided by $A_\mathrm{L}$ only sightly larger than 1 with TT+lowTEB+PP (or TT,TE,EE+lowTEB+PP), and the bestfitting region has a clearly non-zero $\dms \approx 0.012$ ($\dms \approx 0.007$). As expected, PP efficiently breaks the degeneracy between $\dms$ and $A_\mathrm{L}$.

\begin{figure}
\centering
\includegraphics{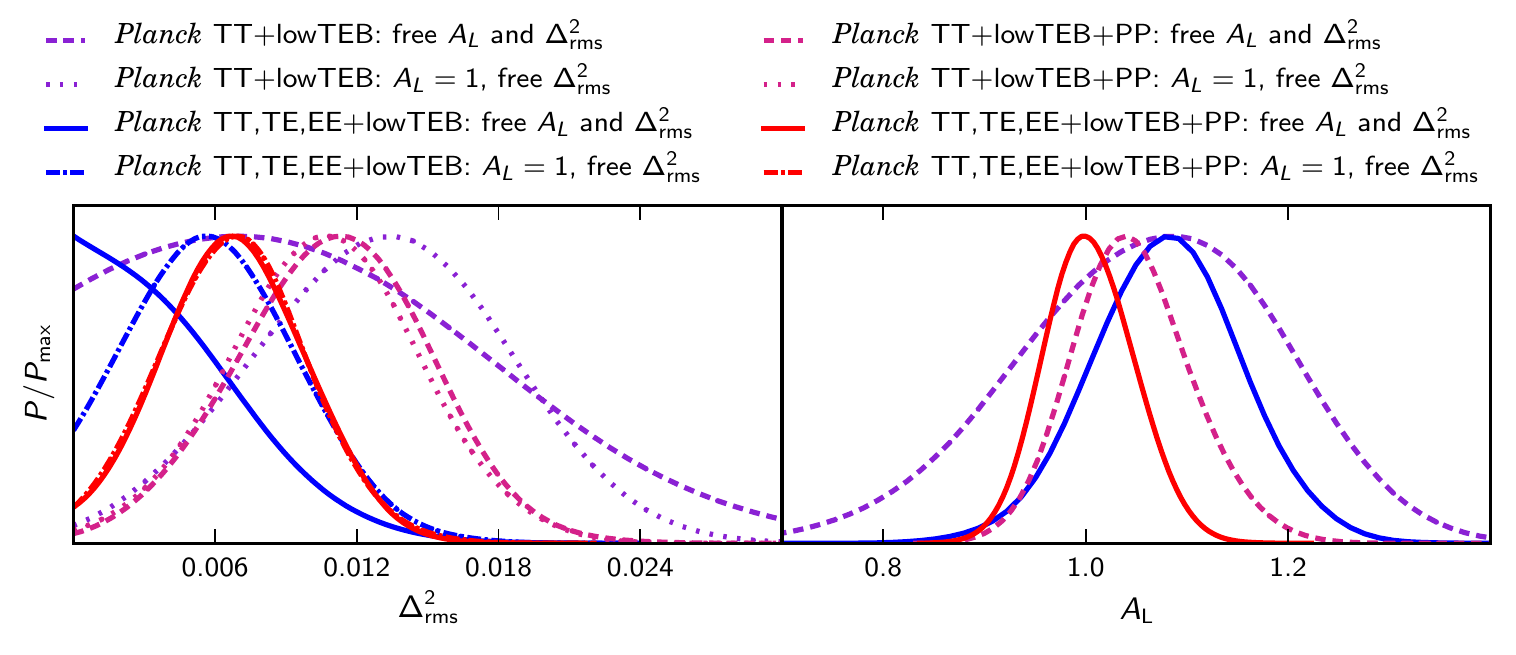}\\
\vspace{-5.5mm}
\caption{
1d marginalized posterior of the CIP variance $\dms$ and lensing amplitude $A_\mathrm{L}$ in two different models ($\Lambda$CDM+$A_\mathrm{L}$+$\dms$ and $\Lambda$CDM+$\dms$ --- the first two cases of table \ref{tab:RealPlanck}) with four combinations of the \Planck\ 2015 data.
\label{fig:planckDeltaAL1d}}
\end{figure}

We repeat the above-described analysis, but fixing $A_\mathrm{L}$ to unity, and compare 1d marginalized posteriors of $\dms$ (and $A_\mathrm{L}$) in Figure~\ref{fig:planckDeltaAL1d} in the cases where both $A_\mathrm{L}$ and $\dms$ are free or $A_\mathrm{L} = 1$ and only $\dms$ is free (in addition to the six $\Lambda$CDM parameters and 15 or 27 nuisance parameters). As the 2d figure suggested, allowing a positive $\dms$ makes smaller values of $A_\mathrm{L}$ favourable (compare to Figure~\ref{fig:planckAL}), since part of the ``extra lensing effect'' is now achieved by CIP. Both with TT and TT,TE,EE plus lowTEB fixing $A_\mathrm{L} = 1$ leads to much larger $\dms$ being favoured than when $A_\mathrm{L}$ is free, since now all the ``extra lensing effect'' in the TT and EE data must be produced by CIP. However, when we add the PP data, the 1d posterior of $\dms$ stays unchanged between the cases of free $A_\mathrm{L}$ and the fixed $A_\mathrm{L}=1$. 

All three cases ($\Lambda$CDM+$A_\mathrm{L}$+$\dms$,  $\Lambda$CDM+$\dms$, and $\Lambda$CDM+$A_\mathrm{L}$)
are quantitatively compared in Table~\ref{tab:RealPlanck} for four combinations of the \Planck\ data. Noteworthy, the posterior mean of $A_\mathrm{L}$ is exactly one with \Planck\ TT,TE,EE+lowTEB+PP data if we allow for compensated isocurvature. Obtaining a big enough lensing-like effect in this case requires $\dms=(6.9^{+2.9}_{-3.2})\times 10^{-3}$ at 68\% CL. Fixing $A_\mathrm{L}$ to one reduces the uncertainty insignificantly: $\dms=(6.9^{+3.0}_{-3.1})\times 10^{-3}$. Our 95\% CL upper bound $\dms < 12 \times 10^{-3}$ with TT,TE,EE+lowTEB+PP coincides with the upper bound $\dms \lesssim 12\times10^{-3}$ obtained from trispectrum in \cite{Grin:2013uya}, whereas using only TT+lowTEB weakens our upper bound by a factor of two.

In \cite{Munoz:2015fdv} lowTEB data was replaced by a prior on the optical depth $\tau=0.068\pm0.019$ and Fisher matrix analysis was employed. Then, with $A_\mathrm{L}=1$, \Planck\ TT gave $\dms < 11\times10^{-3}$ and TT,TE,EE led to $\dms < 5.4\times10^{-3}$ at 68\% CL. Our upper bounds from the full nested sampling are consistent with these, but by a factor of 1.7 larger, since the Fisher-matrix analysis often leads to too optimistic predictions as it ignores the possibly non-gaussian nature of posteriors, in particular any non-linear (banana shaped) degeneracies.

When letting both $A_\mathrm{L}$ and $\dms$ to vary simultaneously, our results are similar to those obtained in \cite{Munoz:2015fdv} by the Fisher matrix analysis using \Planck\ TT or TT,TE,EE plus lowTEB, though the comparison is not straightforward, since also negative values of $\dms$ were allowed in  \cite{Munoz:2015fdv}. When adding the PP data, instead of finding only upper bounds on $\dms$, we get a 2.8$\sigma$ (TT+lowTEB+PP) or 2.3$\sigma$ (TT,TE,EE+lowTEB+PP) ``hint of a detection'' of a non-zero $\dms$, which stays at the similar level if we fix $A_\mathrm{L}=1$; see table \ref{tab:RealPlanck}. (Interestingly, in the \Planck\ ``Cosmological Parameters'' paper \cite{Ade:2015xua}  adding the lensing or BAO data drew all studied one-parameter extensions back to the spatially flat adiabatic $\Lambda$CDM, including the $\Lambda$CDM+$A_\mathrm{L}$ model, which we have also here seen to prefer $A_\mathrm{L}=1$, when the PP data are included. Here we have a model, $\Lambda$CDM+$\dms$, which is not driven to $\dms=0$, when the PP data are taken into account.) However, the preferable values of $\dms$ obtained by the TT+lowTEB+PP data are in slight tension with the 95\% CL upper bound from trispectrum, $\dms\lesssim12\times10^{-3}$  \cite{Grin:2013uya}, and in particular the direct measurements of the variation of baryon fraction in galaxy clusters \cite{Holder:2009gd}, which following Ref.~\cite{Grin:2013uya} leads to $\dms \lesssim 6\times10^{-3}$. The preferred values of $\dms$ obtained by the TT,TE,EE+lowTEB+PP data are more compatible with these ``external'' constraints. This motivates studying in more detail the bestfit parameter combinations and $\chi^2$s for the cases appearing at the last line of table \ref{tab:RealPlanck}, and comparing them to the adiabatic $\Lambda$CDM model.\footnote{Since the bestfit $\chi^2$s in PLA are relatively inaccurate and, on the other hand, as we want to make sure that we use exactly the same settings in the $\Lambda$CDM search as in the extended models, we run the BOBYQA \cite{bobyqa} bestfit search (as implemented in \texttt{CosmoMC}) also for the adiabatic $\Lambda$CDM and $\Lambda$CDM+$A_\mathrm{L}$. (Indeed we find by one point better $\chi^2$ for these models than the PLA values.) For each case we run 12 independent searches starting near the ``bestfit'' found from the \texttt{MultiNest}/\texttt{CosmoMC} chains, allowing for each parameter the 68\% projected range (from \texttt{*.likestats} produced by \texttt{GetDist}), which is broader than the 68\% CL marginalized range (from \texttt{*.margestats}), and using as a proposal matrix the covariance matrix created by \texttt{GetDist} from the full \texttt{MultiNest}/\texttt{CosmoMC} runs. Finally, we pick the best of the 12 ``bestfits''.}

In table \ref{tab:RealPlanckChi2} we use \Planck\ TT,TE,EE+lowTEB+PP data and compare the bestfit $\chi^2$ of three models ( $\Lambda$CDM+$A_\mathrm{L}$+$\dms$,  $\Lambda$CDM+$\dms$, and $\Lambda$CDM+$A_\mathrm{L}$)
to the adiabatic $\Lambda$CDM model. In the bestfit searches we vary, in addition to $A_\mathrm{L}$ and/or $\dms$, all six standard $\Lambda$CDM parameters and the 27 nuisance/foreground parameters. An $A_\mathrm{L}$ much larger than one, which would be required by the TT data, would make the fit to the PP data very bad. Therefore, in the $\Lambda$CDM+$A_\mathrm{L}$ model, PP drives $A_\mathrm{L}$ close to one. Even the moderate bestfit value $A_\mathrm{L}=1.03$ is too large for the PP data and, as seen in the last column, the PP fit becomes by 0.58 worse than in the $\Lambda$CDM case. Overall fit to all the data cannot be improved by $A_\mathrm{L}$, namely $\Delta\chi^2_\mathrm{total}$ is only -0.1. In contrast, a non-zero $\dms$ improves the overall fit considerably, $\Delta\chi^2_\mathrm{total} = -3.6$, and all individual likelihoods show improvement. The $\Lambda$CDM+$\dms$ model fits simultaneously better both the low-$\ell$ and high-$\ell$ data improving also the fit to the PP data, and matching better the calibration and dust priors used in the \Planck\ likelihoods. As already obvious, there is no need to add a free $A_\mathrm{L}$ to the $\Lambda$CDM+$\dms$ model (the improvement would be totally insignificant $\Delta\chi^2\approx -0.1$). As the required value $\dms\sim7\times10^{-3}$ is of the same magnitude as the galaxy cluster upper bounds, we do not strongly endorse CIP as a full solution to the ``lensing anomaly'', but following  Ref.~\cite{Munoz:2015fdv} (which ignored the PP data) point out that CIP could play an important role here, allowing a very good fit to the \Planck\ TT,TE,EE and lowTEB and PP data simultaneously by the addition of only one extra parameter.
\begin{table}
\setlength{\tabcolsep}{0.80mm}
\centering
\footnotesize
\begin{tabular}{|l|r|r|r|}
\hline
 \raisebox{11pt} {}Model & \multicolumn{1}{c|}{$\Lambda$CDM+$A_\mathrm{L}$+$\dms$} &  \multicolumn{1}{c|}{$\Lambda$CDM+$\dms$} & \multicolumn{1}{c|}{$\Lambda$CDM+$A_\mathrm{L}$} \\
 \raisebox{10pt} {} & \multicolumn{1}{c|}{(free $A_\mathrm{L}$ and $\dms$)} &  \multicolumn{1}{c|}{($A_\mathrm{L}=1$, free $\dms$)} & \multicolumn{1}{c|}{($\dms=0$, free $A_\mathrm{L}$)} \\
\hline
\raisebox{10pt} {}Bestfit $1000\dms$ & 7.05 & 7.11 & 0 \\
Bestfit $A_\mathrm{L}$ & 1.02  &  1.00  & 1.03 \\
\raisebox{10pt} {}$\Delta\chi^2_\mathrm{TT,TE,EE}$ & -1.40 &   -1.46 &   -0.25 \\   
\raisebox{9pt} {}$\Delta\chi^2_\mathrm{lowTEB}$ & -1.36 &  -1.09 &  -0.38 \\ 
\raisebox{9pt} {}$\Delta\chi^2_\mathrm{PP}$ & -0.59 &  -0.62 &   +0.58 \\ 
\raisebox{9pt} {}$\Delta\chi^2_\mathrm{prior}$ & -0.37 &  -0.44 &   -0.04 \\ 
\raisebox{9pt} {}$\mathbf{\Delta\chi^2_\mathrm{total}}$ & {\bf -3.72} &  {\bf -3.61} &  {\bf -0.09} \\
\hline
\end{tabular}
\caption{The bestfit $1000\dms$ and $A_\mathrm{L}$ with \Planck\ 2015 TT,TE,EE+lowTEB+PP data in three different models, and the difference of $\chi^2$ of the bestfit compared to the bestfit adiabatic $\Lambda$CDM model. A negative $\Delta\chi^2$ means a better fit to the data than the adiabatic  $\Lambda$CDM model.  
\label{tab:RealPlanckChi2}}
\end{table}

\section{Forecasts for Future Space Missions: LiteBIRD and CORE-M5}
\label{sec:future}

In this section we focus on two concepts of the next-generation satellite mission to measure the polarization of CMB: LiteBIRD \cite{Matsumura:2013aja,Matsumura:2016sri,Grenoble} proposal to JAXA(/NASA) and CORE-M5 \cite{ecoMission,ecoInstrument,DiValentino:2016foa,Finelli:2016cyd,ecoCompSep,ecoCluster,DeZotti:2016qfg} proposal to ESA. We compare their constraining power of $\dms$ and $A_\mathrm{L}$ to simulated \Planck\ data 
 and to the ideal case, where the instrumental noise is zero and the angular resolution infinitely good up to $\ell = 3000$. In the ideal case the only uncertainty will be cosmic variance (assuming $f_\mathrm{sky}=0.7$). Our quantitative results are presented in table \ref{tab:futureDeltaAL} while the figures of this section provide qualitative insight.

\subsection{The studied configurations and simulated data}

For LiteBIRD we use an extended focal plane configuration \cite{Grenoble}, which has 15 frequency channels spanning the range 40\,GHz -- 402\,GHz, with angular resolutions and sensitivities tabulated in \cite{Errard:2015cxa} from where the seven central CMB frequencies, 78\,GHz -- 195\,GHz, are retabulated in \cite{Finelli:2016cyd}. We assume that the eight non-CMB channels are enough for the foreground (synchrotron and polarized dust radiation) removal. Our results present a limit, which can be asymptotically approached with an advanced foreground modeling and cleaning with the help of external experiments \cite{Adam:2015rua,Abazajian:2016yjj}
that may reach much lower and higher frequencies and thus help characterization of the foregrounds. As the main aim of LiteBIRD is detecting the primordial tensor perturbation mode (tensor-to-scalar ratio $r$) via large and medium scale B-mode detection, the angular resolution of LiteBIRD extends only to multipole $\ell \lesssim 1350$. This makes LiteBIRD non-ideal for detecting or constraining compensated isocurvature, which would benefit from the measurement of the smaller scale angular power spectra. Nevertheless we include LiteBIRD here, since it is interesting to compare how an experiment with more sensitivity and in particular with a better angular resolution can improve the constraints.

For CORE-M5 we use the proposed baseline configuration, tabulated, e.g., in \cite{Finelli:2016cyd,DiValentino:2016foa}. It has 2100 detectors and 19 frequency channels spanning the range 60\,GHz -- 600\,GHz. For the forecasts we use six ``conservative CMB channels'', 130\,GHz -- 220\,GHz, which have 956 detectors. The angular resolution stays good up the multipole $\ell \sim 3000$. Again we assume that the other 13 channels (possibly together with external information) are enough for an ideal component separation. 

We create fiducial adiabatic $\Lambda$CDM spectra using the parameters given on the first line of table \ref{tab:fiducial}, which matches the fiducial model employed in \cite{Finelli:2016cyd}. We use \texttt{all\_l\_exact} likelihood of \texttt{CosmoMC}, which is described in \cite{Hamimeche:2008ai} (see also \cite{Knox:1995dq}). This includes the cosmic variance and takes into account that the usable sky fraction $f_\mathrm{sky}$ is not 1, (but instead assumed to be 0.7 in all of our cases). We generate the noise $N_\ell^\mathrm{XX}$ (where XX is TT, EE, or PP) needed in the covariance matrix of the likelihood estimator (and as a direct input for \texttt{CosmoMC}) by the same recipe as in \cite{Finelli:2016cyd,DiValentino:2016foa,Errard:2015cxa}, using for the lensing (PP) noise the quadratic CMB$\times$CMB estimator of \cite{Okamoto:2003zw} (for practical details, see \cite{Errard:2015cxa}). For XX=TT,EE the final $N_\ell^\mathrm{XX}$ will be an inverse-variance weighted sum of the noise sensitivities of the used CMB frequency channels ($\nu$) convolved with a Gaussian beam window function of each channel 
\begin{equation}
\textstyle
N_\ell^\mathrm{XX} = \left[\sum_\nu (1/N_{\ell,\nu}^\mathrm{XX})\right]^{-1},
\end{equation}
where $N_{\ell,\nu}^\mathrm{XX}=\sigma_{\mathrm{X},\nu}^2\exp[\ell(\ell+1)\theta_{\mathrm{FWHM},\nu}/(8\ln2)]$, with $\theta_{\mathrm{FWHM},\nu}$ the full width half maximum beam size of each channel (called ``beam'' in \cite{DiValentino:2016foa} and ``FWHM'' in \cite{Finelli:2016cyd}) and $\sigma_{\mathrm{X},\nu}$ (called $\Delta T$ for X=T and $\Delta P$ for X=E in the tables of \cite{Finelli:2016cyd,DiValentino:2016foa}) the noise sensitivity, which should be converted from $\mu$K$\cdot$arcmin to $\mu$K$\cdot$steradian before inserting into this formula.

Since it would not be feasible or reasonable to reproduce all the artefacts and ``anomalies'' of the real \Planck\ data to our simulated LiteBIRD, CORE-M5, and cosmic variance limited data, we take the opposite approach in order to be able to compare their sensitivities to \Planck. Namely, we create fiducial adiabatic $\Lambda$CDM data with the pipeline described above also for \Planck, using bluebook values \cite{PlanckBlueBook}  of the 100, 143, and 217\,GHz High Frequency Instrument (HFI) channels for the angular resolution and noise sensitivity (the latter divided by $\sqrt{2}$ to take into account that HFI operated approximatively twice as long as the nominal mission was planned to last). In figures we denote this data set by \Planck\ (sim.).

\subsection{Constraints on $\dms$}

\begin{table}
\setlength{\tabcolsep}{0.9mm}
\footnotesize
\begin{tabular}{|l|rr|rr||rr||rr|}
\hline
\raisebox{11pt} {}Model & \multicolumn{4}{c||}{$\Lambda$CDM+$A_\mathrm{L}$+$\dms$} &  \multicolumn{2}{c||}{$\Lambda$CDM+$\dms$} & \multicolumn{2}{c|}{$\Lambda$CDM+$A_\mathrm{L}$}\\
 \raisebox{11pt} {}   & \multicolumn{4}{c||}{$\phantom{1}$\hspace{0.75cm}(free $A_\mathrm{L}$ and $\dms$)\hspace{0.75cm}$\phantom{1}$} &  \multicolumn{2}{c||}{($A_\mathrm{L}=1$, free $\dms$)} & \multicolumn{2}{c|}{($\dms=0$, free $A_\mathrm{L}$)}\\
\hline
 \raisebox{11pt} {}Parameter   &  \multicolumn{2}{c|}{$\phantom{1}$\hspace{4mm}$1000\dms$\hspace{4mm}$\phantom{1}$} &  \multicolumn{2}{c||}{$\phantom{1}$\hspace{4mm}$100A_\mathrm{L}$\hspace{4mm}$\phantom{1}$} &  \multicolumn{2}{c||}{$\phantom{1}$\hspace{4mm}$1000\dms$\hspace{4mm}$\phantom{1}$} &  \multicolumn{2}{c|}{$\phantom{1}$\hspace{4mm}$100A_\mathrm{L}$\hspace{4mm}$\phantom{1}$} \\
    & 68\% CL & 95\%  CL & mean & $\sigma$ &  68\% CL & 95\% CL & mean & $\sigma$ \\
\hline
\Planck\ (sim.) TT,TE,EE &      $<$ 4.28 &      $<$ 8.85 &     95.30 &       4.62 &      $<$  2.47 &     $<$   5.04 &     98.87 &      3.81  \\
LiteBIRD TT,TE,EE &       $<$ 4.73 &      $<$  9.48 &     90.92 &      7.80 &      $<$  2.90 &      $<$  6.16 &     95.64 &      7.14  \\
CORE-M5 TT,TE,EE &  $<$  1.27 &      $<$  2.67 &     98.83 &      1.53 &   $<$   0.90 &      $<$  1.89 &     99.74 &      1.31  \\
{\bf CORE-M5 TT,TE,EE,PP} &      $<$  0.66 &     $<$   {\bf 1.42} &     99.38 &      1.21 &      $<$  0.60 &      $<$  {\bf 1.35} &     99.55 &      1.18  \\
Ideal TT,TE,EE lmax=3000 &     $<$   0.75 &      $<$  1.61 &     99.41 &      1.19 &     $<$   0.63 &      $<$  1.32 &     99.97 &      1.10  \\
\hline
\end{tabular}
\caption{The posterior 68\% CL and 95\% CL upper bounds on $1000\dms$, and the posterior mean value and standard deviation for $100A_\mathrm{L}$ in three different models (first $\Lambda$CDM+$A_\mathrm{L}$+$\dms$, second  $\Lambda$CDM+$\dms$, third  $\Lambda$CDM+$A_\mathrm{L}$) using simulated adiabatic $\Lambda$CDM ($\dms=0$, $A_\mathrm{L}=1$) \Planck\ data, or three different configurations of future space missions or, as the last one, an ideal case, in which the only ``noise'' is cosmic variance with $f_\mathrm{sky}=0.7$.
\label{tab:futureDeltaAL}}
\end{table}

We start by fitting the eight parameter $\Lambda$CDM+$A_\mathrm{L}$+$\dms$ model in figure \ref{fig:futureDeltaAL2d} and the first column of table  \ref{tab:futureDeltaAL}.  While LiteBIRD performs almost as well as (simulated) \Planck\ data in constraining $\dms$, the LiteBIRD constraints on $A_\mathrm{L}$ are much weaker. CORE-M5 TT,TE,EE performs 3.3 times better for $\dms$ and 3.0 times better for $A_\mathrm{L}$ than \Planck. In \cite{Finelli:2016cyd} CORE-M5 TT,TE,EE was found to almost reach the cosmic variance limited ideal case for most of CDM isocurvature models studied, but here we find that for CIP the CORE TT,TE,EE constraint is 1.7 times weaker than in the ideal case. (The $A_\mathrm{L}$ constraint is only 1.3 times weaker.) However, adding PP improves the CORE-M5 95\% CL constraint to $\dms< 1.4\times10^{-3}$, which is 13\% better than in the ideal case with TT,TE,EE.

In addition to the previous cases, we show 1d marginalized posteriors for $\dms$ (and $A_\mathrm{L}$) in figures \ref{fig:futureDeltaAL1d} and \ref{fig:COREDeltaAL1d} also for the $\Lambda$CDM+$\dms$ model (the second column of table \ref{tab:futureDeltaAL}). For LiteBIRD and \Planck, due to the long $\dms$--$A_\mathrm{L}$ degeneracy line the upper bound on $\dms$ becomes much tighter (by a factor of 1.5 and 1.8, respectively) if $A_\mathrm{L}$ is fixed to one. As the degeneracy is much reduced by CORE TT,TE,EE, its results for $\dms$ differ less between the free and fixed $A_\mathrm{L}$ cases (by a factor of 1.4). The addition of PP in practice breaks the degeneracy (see the green curves in figures \ref{fig:futureDeltaAL2d} and \ref{fig:COREDeltaAL1d} and the line in bold in table \ref{tab:futureDeltaAL}) and thus decreases the difference to the factor of 1.42/1.35=1.05, which is much less than 1.61/1.32=1.2 of the ideal TT,TE,EE case.

Due to the inability of TT,TE,EE data to distinguish between $\dms$ and $A_\mathrm{L}$, it is fair to quote as our main results the weaker constraints obtained in the $\Lambda$CDM+$A_\mathrm{L}$+$\dms$ model rather than advocating the tighter upper bounds found in the $\Lambda$CDM+$\dms$ and $\Lambda$CDM+$A_\mathrm{L}$ models. However, for CORE-M5 with the lensing data included (highlighted by bold face in table \ref{tab:futureDeltaAL}) the results are identical up to one decimal place: $\dms < 1.4\times10^{-3}$ at 95 \% CL and $\sigma(A_\mathrm{L}) =1.2\times10^{-2}$, and we quote these as our best case forecasts. The predicted sensitivity of the power spectra of CORE-M5 to CIP is 9 times better than the current upper bounds for $\dms$ from trispectrum \cite{Grin:2013uya}, but two orders of magnitude worse than may be obtained from a cosmic variance limited trispectrum measurement \cite{He:2015msa}.
\afterpage{
\begin{figure}
\centering
\includegraphics[width=0.425\textwidth]{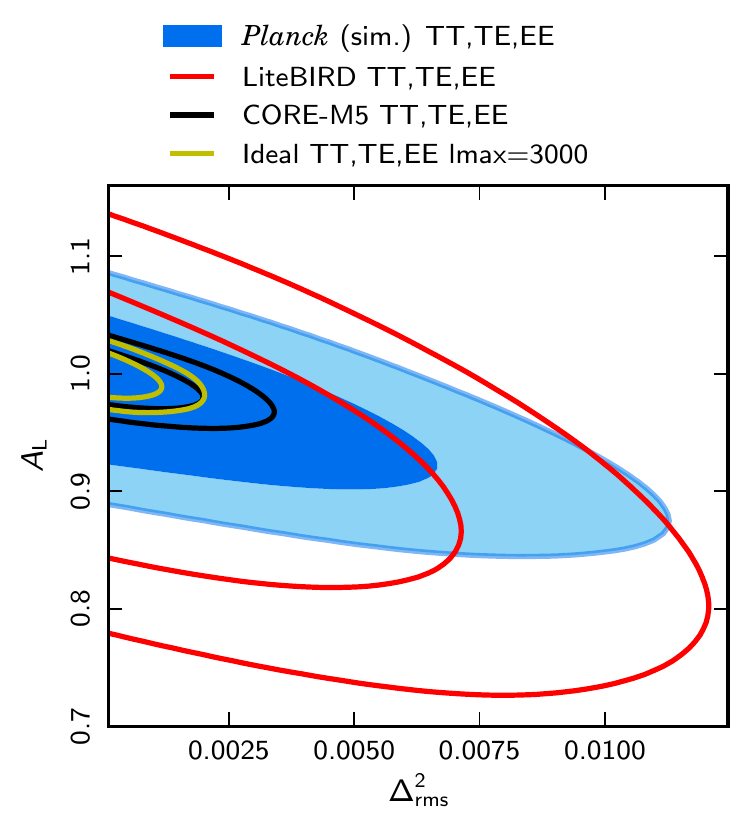}
\includegraphics[width=0.425\textwidth]{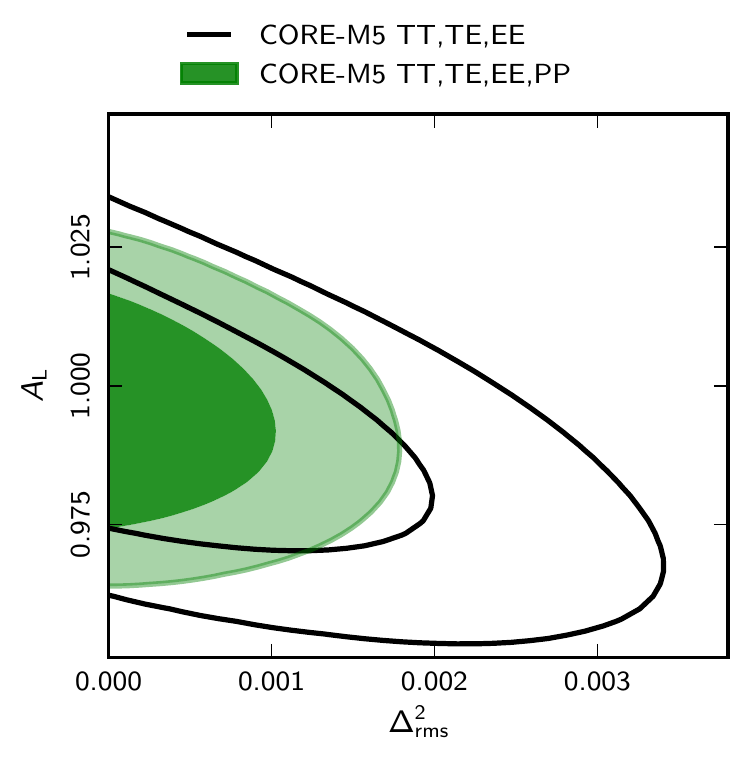}\\
\vspace{-5.5mm}
\caption{
2d marginalized 68\% and 95\% CL posterior regions of $A_\mathrm{L}$ and $\dms$, when the simulated data are based on the pure adiabatic $\Lambda$CDM model with $A_\mathrm{L}=1$. The zoom-up in the right panel shows how much the inclusion of CORE-M5 lensing data (PP) may improve over CORE-M5 TT,TE,EE data.
\label{fig:futureDeltaAL2d}}
\end{figure}
\begin{figure}
\centering
\includegraphics[width=0.85\textwidth]{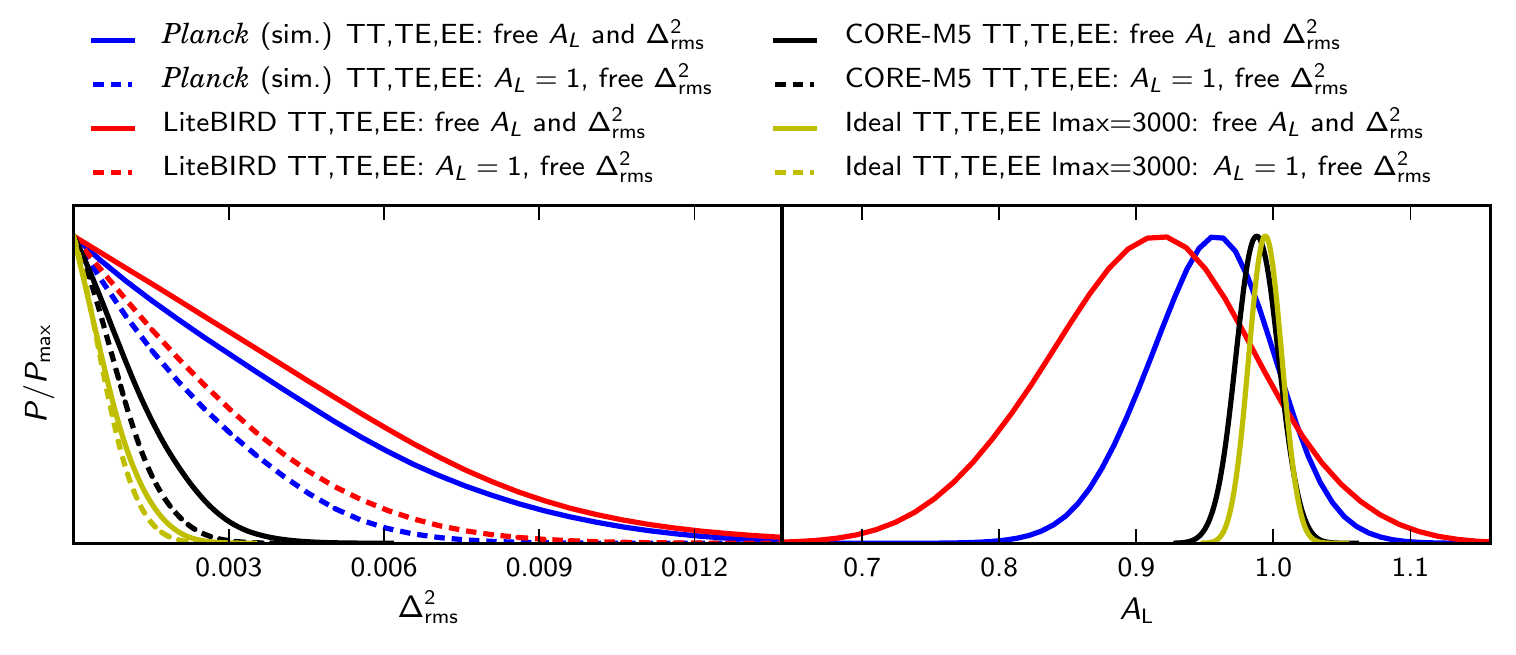}\\
\vspace{-5.5mm}
\caption{
1d marginalized posterior of the CIP variance $\dms$ and lensing amplitude $A_\mathrm{L}$ in two different models, $\Lambda$CDM+$A_\mathrm{L}$+$\dms$ (solid lines) and $\Lambda$CDM+$\dms$ (dashed lines) --- the first two cases of table \ref{tab:futureDeltaAL}. The simulated data are based on the pure adiabatic $\Lambda$CDM model with $A_\mathrm{L}=1$.
\label{fig:futureDeltaAL1d}}
\end{figure}
\begin{figure}
\centering
\includegraphics[width=0.85\textwidth]{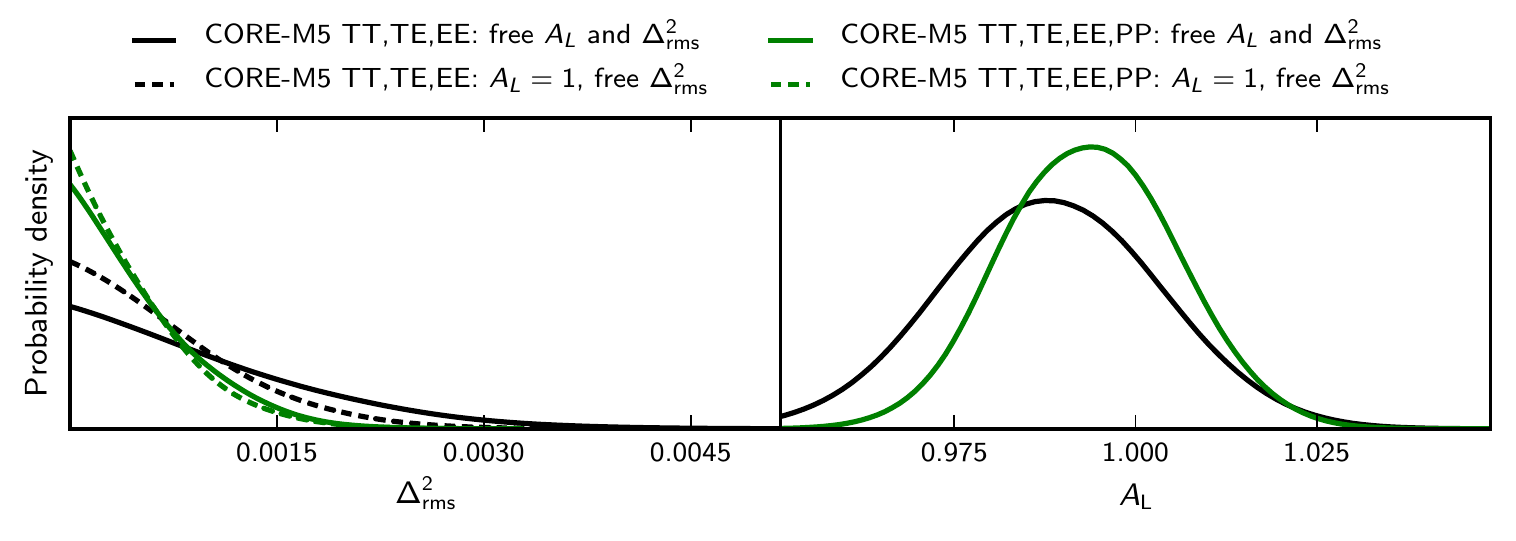}\\
\vspace{-5.5mm}
\caption{
The same as figure \ref{fig:futureDeltaAL1d}, but now comparing CORE-M5 with TT,TE,EE (black lines) to the case where also the lensing data (PP) are included (green lines).
\label{fig:COREDeltaAL1d}}
\vspace{-2.5mm}
\end{figure}}

\subsection{Constraints on $A_\mathrm{L}$}

\begin{figure}
\centering
\includegraphics[width=0.85\textwidth]{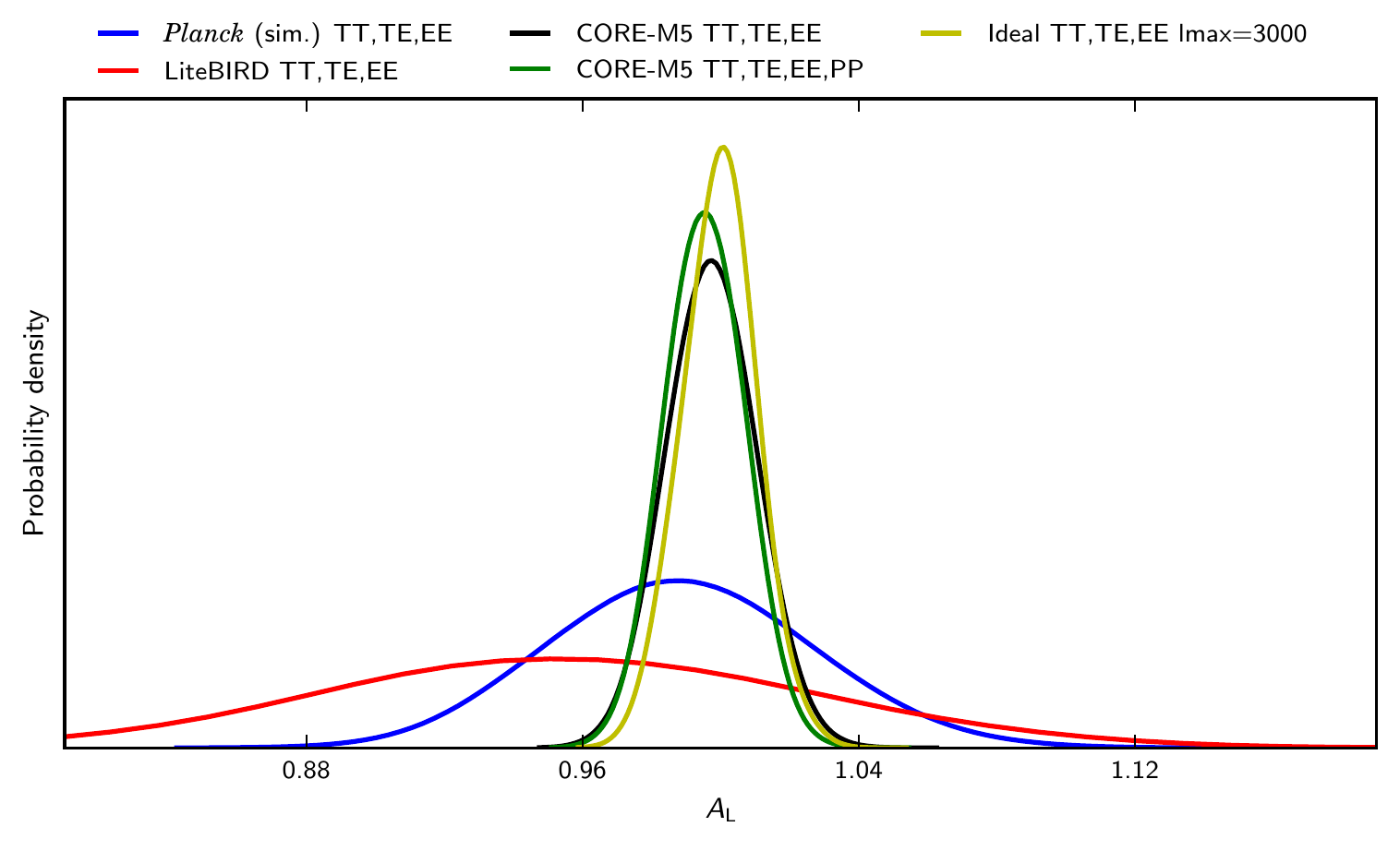}\\
\vspace{-5.5mm}
\caption{
The posterior of the phenomenological lensing amplitude $A_\mathrm{L}$, when the simulated data are based on the pure adiabatic $\Lambda$CDM model with $A_\mathrm{L}=1$ and the fitted model is $\Lambda$CDM+$A_\mathrm{L}$ (the last model of table \ref{tab:futureDeltaAL}).
\label{fig:futureAL}} 
\end{figure}

If, in the future, the trispectrum, the galaxy cluster, or other ``external'' upper bounds turn out order(s) of magnitude stronger than the forecasted sensitivity of the power spectra to $\dms$ presented in the previous subsection, then it is safe to assume $\dms=0$ when studying $A_\mathrm{L}$. In figure \ref{fig:futureAL} we present forecasts for the $\Lambda$CDM+$A_\mathrm{L}$ model, employing simulated adiabatic $\Lambda$CDM data with $A_\mathrm{L}=1$. Now the uncertainty of the determination of $A_\mathrm{L}$ shrinks compared to the $\Lambda$CDM+$A_\mathrm{L}$+$\dms$ case, as quantitatively confirmed by comparing the last and first models of table \ref{tab:futureDeltaAL}. However, this shrinkage is very moderate: even for the simulated \Planck\ data, where the effect is largest, $\sigma(A_\mathrm{L})$ in the $\Lambda$CDM+$A_\mathrm{L}$ model is only 18\% smaller than in the $\Lambda$CDM+$A_\mathrm{L}$+$\dms$ model.

However, assuming the $\Lambda$CDM+$A_\mathrm{L}$+$\dms$ model and trying to determine  $A_\mathrm{L}$ from 1d marginalized posterior (where one integrates over the $\dms$ direction, e.g., in figure \ref{fig:futureDeltaAL2d}) introduces a significant bias toward smaller values of $A_\mathrm{L}$ than the input value $A_\mathrm{L}=1$. The simulated \Planck\ data give for the $\Lambda$CDM+$A_\mathrm{L}$+$\dms$ case $A_\mathrm{L} = 0.953$ and for the $\Lambda$CDM+$A_\mathrm{L}$ case $A_\mathrm{L} = 0.989$. As obvious from table \ref{tab:futureDeltaAL} and the right panel of figure \ref{fig:COREDeltaAL1d}, CORE-M5 (in particular with the PP data) leads to almost symmetrical posterior of $A_\mathrm{L}$ about 1 due to much reduced $\dms$--$A_\mathrm{L}$ degeneracy.

CORE-M5 determines $A_\mathrm{L}$ six times better than LiteBIRD, and three times better than the (optimistically) simulated \Planck\ data. With the TT,TE,EE data, CORE-M5 falls only 20\% short from the ideal cosmic variance limited case. With the help of PP this reduces to less than 10\%. 

Finally, we point out that if the value $A_\mathrm{L}\sim1.22$ favoured by the real \Planck\ TT data was ``the true value'', then CORE-M5 would detect this at $\sim\!\!14$--$19\sigma$ level.\footnote{With the simulated \Planck\ data $A_\mathrm{L}\sim1.22$ is about $5\sigma$ away from one. Part of the difference of $\sigma(A_\mathrm{L})$ between the simulated and real \Planck\ data comes from the much better determination of $\tau$ by the simulated \Planck\ data, and the rest comes from the fact that in the simulated \Planck\ we have also the TE and EE components created from the same $\Lambda$CDM model as the TT component. Furthermore, in the simulated TT data there is no ``lack of power'' at low multipoles as there is in the real lowTEB data.} While the $2.2\sigma$ deviation of \Planck\ TT from the $\Lambda$CDM value is only a minor hint toward new physics or merely unresolved systematics/foregrounds, CORE-M5 would be able to unambiguously detect a deviation of this magnitude from the $\Lambda$CDM model.

\section{Conclusions}

We have studied compensated isocurvature perturbations between baryons and cold dark matter that can be described at observable scales (multipoles 2--3000 in our case) as a small anisotropy of the baryon (and CDM) density of the Universe, such that $\delta\rho_b(\hat n) = -\delta\rho_c(\hat n)$ in the direction $\hat n$ on the sky. Defining $\Delta(\hat n) = [\rho_b(\hat n) - \rho_b] / \rho_b$, where $\rho_b$ is the average baryon density and treating $\Delta(\hat n)$ as a Gaussian random variable with zero mean \cite{Munoz:2015fdv}, the compensated isocurvature perturbations can be characterized by a single parameter, the variance $\dms \equiv \langle |\Delta(\hat n)|^2 \rangle$.

Replacing the idealistic Fisher matrix analysis of \cite{Munoz:2015fdv} by a full nested sampling where we vary six $\Lambda$CDM parameters, as well as 15 or 27 nuisance/foreground parameters of the \Planck\ likelihoods, $\dms$, and optionally the phenomenological lensing amplitude parameter $A_\mathrm{L}$, we have derived \Planck\ 2015 power spectra based constraints on $\dms$, which are of the same order of magnitude as the current trispectrum based constraints $\dms \lesssim 12\times10^{-3}$ at 95\% CL. This value corresponds to the baryon and CDM density isocurvature power that is order(s) of magnitude larger than the primordial curvature perturbation power, which is about $\mathcal{P_R}\approx2.1\times 10^{-9}$. This is due to the fact that, in the case of compensated isocurvature, the total matter isocurvature mode is zero, and hence the observational signal in the CMB power spectra is weak.

We have shown that a non-zero $\dms$ significantly improves the simultaneous fit to the \Planck\ 2015 temperature,\footnote{Previously the ability of $\dms$ to improve the fit to the \Planck\ 2015 temperature data was noticed in \cite{Munoz:2015fdv}. However, there are many other extensions of $\Lambda$CDM model that can improve the fit to the temperature data without needing $A_\mathrm{L}\ne1$ or at least can bring $A_\mathrm{L}$ closer to its $\Lambda$CDM value $A_\mathrm{L}=1$. However, it has turned out difficult to devise models that would not then fit worse the lensing data. We point out in this paper that the compensated isocurvature model is capable for a good joint fit. For example, a two-parameter modified gravity (MG) model studied in \cite{DiValentino:2015bja} is $2\sigma$ favoured by \Planck\ TT and many other datasets, but for the PP data Ref.~\cite{DiValentino:2015bja} concludes: ``However it also important to stress that when the CMB lensing likelihood is included in the analysis the statistical significance for MG simply vanishes.''  Similarly, in \cite{Valiviita:2015dfa} an interaction between dark matter and (phantom) dark energy is almost $3\sigma$ favoured  by the \Planck\ 2013 TT + BAO data in a phenomenological one-parameter model, but adding the PP data reduces this to $1.5\sigma$.} polarization, and lensing data: $\dms \approx 7\times 10^{-3}$  helps to reduce the bestfit $\chi^2$ by 3.6 compared to the spatially flat adiabatic $\Lambda$CDM model.  Since the effect of a non-zero $\dms$ on the temperature and polarization power spectra is similar to the effect of an enhanced CMB lensing via the phenomenological lensing parameter $A_\mathrm{L}>1$, there is no need for $A_\mathrm{L} \ne 1$. Unlike  $A_\mathrm{L} \ne 1$, a non-zero $\dms$ leaves the lensing potential power spectrum $C_\ell^{\phi\phi}$ almost unchanged, and thus does not spoil the fit to the $C_\ell^{\phi\phi}$ data as does $A_\mathrm{L} \simeq 1.22$, which is favoured by the \Planck\ high-multipole temperature data (if $\dms$ is kept zero). Therefore, compensated isocurvature provides an example of a \emph{simple} model, which is capable of reducing the \Planck\ lensing anomaly significantly and fitting well \emph{simultaneously} the high-$\ell$ temperature and lensing potential reconstruction data. 

After discussing the \Planck\ results, the $\dms$--$A_\mathrm{L}$ degeneracy (if only temperature and polarization data are used), and illuminating how the lensing data can break it, we have presented forecasts for the future CMB space missions: the LiteBIRD \cite{Matsumura:2013aja,Matsumura:2016sri,Grenoble} proposal to JAXA/NASA and CORE-M5 \cite{ecoMission,ecoInstrument,DiValentino:2016foa,Finelli:2016cyd,ecoCompSep,ecoCluster,DeZotti:2016qfg} proposal to ESA. Since LiteBIRD is optimized for large and medium scale polarization B-mode detection, its angular resolution degrades above multipole $\ell \sim 1350$. For the detection of the lensing-like effect of CIP higher multipoles would be beneficial. Hence LiteBIRD, even in its extended focal plane configuration \cite{Errard:2015cxa,Finelli:2016cyd} studied here, is not expected to improve over \Planck, what comes to the constraints on $\dms$ or  $A_\mathrm{L}$. In contrast, CORE-M5 can reach multipoles up to $\ell \sim 3000$, and has also a capability for accurate lensing potential reconstruction at high multipoles. Therefore CORE-M5 will be able to distinguish between $\dms$ and  $A_\mathrm{L}$ and will constrain them exquisitely even if both are varied simultaneously --- almost as well as an ideal, cosmic variance limited, experiment. Employing simulated adiabatic $\Lambda$CDM data with $A_\mathrm{L}=1$ our forecast for CORE-M5 is $\dms < 1.4\times10^{-3}$ at 95\% CL, which is nine times stronger than the current trispectrum based constraints and six times stronger than we obtain from simulated \Planck\ data with the same pipeline.  

As a side product of our analysis we have obtained forecasts for the determination accuracy of $A_\mathrm{L}$: we have found $\sigma(A_\mathrm{L}) = 0.038, 0.071, 0.012, 0.011$ for the simulated \Planck, LiteBIRD, CORE-M5, and cosmic variance limited data, respectively. CORE-M5 represents huge potential for detecting or ruling out the deviations from the adiabatic $\Lambda$CDM model. If the true lensing amplitude was the one favoured by the \Planck\ temperature data, $A_\mathrm{L}\simeq1.22$, CORE-M5 would be able to detect this at 14--19$\sigma$ level.\\

\acknowledgments

I acknowledge funding from the Finnish Cultural Foundation and thank CSC -- the IT Center for Science Ltd. (Finland) for computational resources.  This work was supported in part by the Academy of Finland grants 257989 and 295113. The results with the real \Planck\ data are based on observations obtained with \Planck\ (\url{http://www.esa.int/Planck}), an ESA science mission with instruments and contributions directly funded by ESA Member States, NASA, and Canada.

\bibliographystyle{JHEP}
\bibliography{CompIsoc}



\end{document}